\documentclass[twoside,twocolumn,10pt]{article}
\usepackage[utf8x]{inputenc}

\usepackage[widespace,poorman]{fourier} 
\usepackage{microtype}

\usepackage[table]{xcolor}
\usepackage{graphicx}
\usepackage{subfigure}
\usepackage{float}
\usepackage{amsmath}
\usepackage{amssymb}
\usepackage{tikz}
\usepackage{pgfplots}
\usepackage{enumerate}
\usepackage{fancybox}
\usetikzlibrary{arrows,snakes,shapes}
\usepackage{epstopdf}
\usepackage{hyperref}
\usepackage{cases}
\usepackage{array}
\usepackage{multirow}
\usepackage{booktabs}
\usepackage{setspace}
\usepackage{appendix}
\usepackage{geometry}
\usepackage[T1]{fontenc}

\usepackage{fancybox}
\usepackage{fancyhdr}
\hypersetup{
backref=true,            
pagebackref=true,    
hyperindex=true,      
colorlinks=true,        
breaklinks=true,       
urlcolor= orange,        
linkcolor= blue,        
citecolor=red,
bookmarks=true,       
bookmarksopen=false,  
}

\newcommand{\EV}[1]{\mathbb{E}\left[#1\right]}
\newcommand{\STD}[1]{\text{STD}\left[#1\right]}

\usepackage[font=small,labelfont={bf,sf}]{caption}
\usepackage{paralist}
\usepackage{multicol}

\usepackage{abstract}
	
\usepackage{titlesec}
\titleformat{\section}[block]{\center\large\scshape\bfseries\sffamily}{\thesection.}{0.6em}{}
\titleformat{\subsection}[block]{\normalsize\bfseries\sffamily}{\thesubsection}{0.6em}{}
\titleformat{\subsubsection}[block]{\normalsize\itshape}{\thesubsubsection}{0.6em}{}

\fancyheadoffset[LE,RO]{0cm}

\fancypagestyle{plain}{%

\fancyhf{} 
}

\title{\Large\vspace{-20mm}%
	\sffamily
	\textbf{Prediction of the dynamic oscillation threshold in a clarinet model with a linearly increasing blowing pressure : Influence of noise}
	}	
\author{%
	\large
	\textsc{B. Bergeot$^{a,}$\footnote{Corresponding author, \texttt{baptiste.bergeot@univ-lemans.fr}} , A. Almeida$^{a}$, C. Vergez$^{b}$, B. Gazengel$^{a}$}}

\date{{\small \textit{$^{a}$LUNAM Universit\'{e}, Universit\'{e} du Maine, UMR CNRS 6613, Laboratoire d’Acoustique, Avenue Olivier Messiaen, 72085 Le Mans Cedex 9, France}}\\
{\small \textit{$^{b}$Laboratoire de M\'{e}canique et Acoustique (LMA, CNRS UPR7051), 31 Chemin Joseph Aiguier, 13402 Marseille Cedex 20, France}}}

\begin{document}

\twocolumn[
    \maketitle
    \hrulefill
\begin{onecolabstract}
\noindent This paper presents an analysis of the effects of noise and precision on a simplified model of the clarinet driven by a variable control parameter.

When the control parameter is varied the clarinet model undergoes a dynamic bifurcation. A consequence of this is the phenomenon of bifurcation delay: the bifurcation point is shifted from the \textit{static} oscillation threshold to an higher value called \textit{dynamic} oscillation threshold.

In a previous work \cite{BergeotNLD2012}, the dynamic oscillation threshold is obtained analytically. In the present article, the sensitivity of the dynamic threshold on precision is analyzed as a stochastic variable introduced in the model. A new theoretical expression is given for the dynamic thresholds in presence of the stochastic variable, providing a fair prediction of the thresholds found in finite-precision simulations. These dynamic thresholds are found to depend on the increase rate and are independent on the initial value of the parameter, both in simulations and in theory.
\end{onecolabstract}
   \hrulefill
\vspace{0.7cm}]
{
  \renewcommand{\thefootnote}%
    {\fnsymbol{footnote}}
  \footnotetext[1]{{Corresponding author, \texttt{baptiste.bergeot@univ-lemans.fr}}
}


\begin{table}[t]
\centering
{\small
\rowcolors{1}{white}{gray!25}
\caption{Table of notation. All quantities are dimensionless.}
\label{tab:TabOfNot}       
\begin{tabular}{|p{0.8cm}|p{6.6cm}|}
\hline
\multicolumn{2}{ |c| }{\textbf{Table of Notation}}\\ \hline
$G$ & iterative function \\
$\gamma$ & musician mouth pressure (control parameter) \\
$\zeta$ & control parameter related to the opening of the embouchure at rest \\
$p^+$ & outgoing wave \\
$p^-$ & incoming wave \\
$p^{+*}$ & non-oscillating static regime of $p^+$ (fixed points of the function $G$) \\
$\phi$ & invariant curve \\
$w$ & difference between $p^+$ and $\phi$ \\
$\epsilon$  &increase rate of the parameter $\gamma$ \\
$\sigma$  & level of the white noise  \\
$\gamma_{st}$ & static oscillation threshold \\
$\gamma_{dt}$ & dynamic oscillation threshold \\
$\gamma_{dt}^{th}$ & theoretical estimation of the dynamic oscillation threshold of the clarinet model without noise or in "deterministic" situation\\
$\hat{\gamma}_{dt}^{th}$ & theoretical estimation of the dynamic oscillation threshold of the noisy clarinet model in "sweep-dominant" situation  \\
$\Gamma_{dt}^{th}$ & general theoretical estimation of the dynamic oscillation threshold of the noisy clarinet model, both for a "sweep-dominant regime" or a "deterministic regime" \\
$\gamma_{dt}^{num}$  & dynamic oscillation threshold calculated on numerical simulations \\
\hline
\end{tabular}
}
\end{table}

\section{Introduction}
\label{sec:introduction}

In classical (or \textit{static}) bifurcation theory, all the parameters are constant, including the bifurcation parameter. The \textit{dynamic} bifurcation theory focuses on systems where the bifurcation parameter is varying slowly over time. For a given system, the location of the bifurcation can be significantly different in the latter case.

A simple illustration is the flip bifurcation undergone by many one-dimensional discrete time nonlinear systems (among which the well known logistic map \cite{KuzAppBif2004} or a clarinet model \cite{Maga1986,NonLin_Tail_2010}). When the bifurcation parameter is constant, the \textit{static} bifurcation diagram summarizes the behavior of the system around the bifurcation : below the critical value of the parameter, the fixed point is stable (thus attractive), and above the critical value of the parameter the fixed point is unstable (thus repulsive) whereas a 2-valued cycle, born at the bifurcation, is stable (thus attractive). When the bifurcation parameter is varied over time, a \textit{bifurcation delay} may appear : when the \textit{static} bifurcation point is passed, the orbit remains in the neighborhood of the branch of the fixed points. After a certain time, the \textit{dynamic} bifurcation point is reached: the system escapes from the branch of the fixed points and moves abruptly to the 2-valued cycle. This behavior may be depicted in a \textit{dynamic bifurcation diagram}. Fruchard and Schäfke~\cite{Fruchard2007} published an overview of the problem of bifurcation delay.

A previous article by the authors~\cite{BergeotNLD2012} analyzed the behavior of a simplified model of a clarinet when one of its control parameters (the blowing pressure) increases slowly linearly with time. Oscillations corresponding to the production of sound start at a much higher threshold than the one obtained in a static parameter case (i.e. higher than \textit{static} bifurcation point of the system). The dynamic threshold (i.e. the \textit{dynamic} bifurcation point) was described by an analytical expression, predicting that it does not depend on the increase rate of the blowing pressure (within the limits of the theory, i.e.~slow enough increases), but that it is very sensitive to the starting value of the linear increase. This is a known behavior of such kind of system which time-varying parameter, shown by Baesens~\cite{Baesens1991} and, in the framework of nonstandard analysis, by Fruchard~\cite{FruchInstFour1992}.

These results are reproduced by simulations of the model, but only when very high precisions are used in the simulations. Running the simulation with the normal double-precision of a CPU results in much lower thresholds, although higher than the static ones.

The problem of the precision had already been mentioned in the seminal article~\cite{ChasCanard} \textit{Chasse au canard} (\textit{Duck Hunting} in english).The \textit{canard} phenomenon have similarities with the bifurcation delay. However, it can also appear in static situations: if the control parameter is higher than the static bifurcation point a stable limit cycle appears but, in particular cases, a delay can be observed in the limit cycle itself (see \cite{ChasCanard,Fruchard2007} for \textit{canards} of forced Van der Pol equation). The shape of the resulting \textit{canard} cycle in the phase space resembles that of a duck . This phenomenon can only exist in a very narrow interval of the parameter. Consequently, numerical simulations have to be performed with high precision and it was impossible in the beginning of the 80ies.

For the dynamic bifurcation, in contrast with the theory and simulations using high precision, when numerical simulations are running with finite precision, the dynamic threshold depends on the parameter increase rate, but doesn't depend on the starting value of the parameter. These properties have been observed on numerical simulations of the logistic map by Kapral and Mandel~\cite{Kapral1985} and in~\cite{BergeotNLD2012} in the case of a clarinet model.

To explain this discrepancy, round-off errors of the computer must be taken into account. In general this is done by introducing an ad hoc additive white noise in the model. For continuous-time systems we can cite Benoît~\cite{DynBifLumEben1990} and more recently Berglund and Gentz~\cite{BGNonlinearity2002,BGProb2002}. For discrete-time systems Baesens~\cite{Baesens1991,Baesens1991Noise} propose a general method which is followed in the present paper.

Therefore, the aim of the present article is to formulate analytically an estimation of the dynamic bifurcation threshold in simulations performed with finite precision. The effect of finite numerical precision in simulations is modeled as an ad hoc additive white noise with uniform distribution. This hypothesis is tested in section \ref{s:pr_noise}. In section \ref{sec:SeDynNoTh}, a mathematical relation is derived for the behavior of the model affected with noise. The resulting theoretical expression of the dynamic oscillation threshold is compared to numerical simulations and its range of validity is discussed. The clarinet model and major results from~\cite{BergeotNLD2012} are first briefly recalled in section \ref{sec:recall}.

A table of notation is provided in table~\ref{tab:TabOfNot}.
 
\section{Dynamic oscillation threshold of the clarinet model without noise and problem statement}
\label{sec:recall}

\subsection{Clarinet model}

The instrument is divided into two functional elements: the exciter and the resonator. The exciter of the clarinet is the reed-mouthpiece system described by a nonlinear characteristics relating, by the Bernoulli equation, the instantaneous values of the flow $u(t)$ across the reed entrance to the pressure difference $\Delta p(t)=p_m(t)-p(t)$ between the  mouth of the musician  and the clarinet mouthpiece \cite{HirschAal1990,MOMIchap7}. The reed is simplified into an ideal spring without damping or inertia. The resonator is approximated by a straight cylinder,  described by its reflection function $r(t)$. Considering that the resonator is a perfect cylinder in which the dispersion is ignored and the losses are assumed to be frequency independent \cite{Maga1986,MechOfMusInst}. The reflection function $r(t)$ becomes a simple delay with sign inversion (multiplied by an loss parameter $\lambda$) and is written:
 
\begin{equation}
r(t) = - \lambda \delta(t - \tau),
\label{eq:DiracR}
\end{equation}
where $\tau =2 L /c$ is the travel time for waves to propagate to the end of the resonator of length $L$ at speed $c$ and to return to the input.

The loss parameter $\lambda$ takes into account the visco-thermal losses along the resonator, 
which at low frequencies are dominant over the radiation losses. It can be approximated by the expression:

\begin{equation}
\lambda=e^{-2\alpha L},
\label{lambda}
\end{equation}
where $\alpha$ is the damping factor \cite{KeefeJASA1984}:
\begin{equation}
\alpha \approx 3 \cdot 10^{-5} \sqrt{f}/R.
\label{alpha}
\end{equation}

$R$ is the resonator radius and $f$ is chosen to be the fundamental playing frequency. A realistic value of the loss parameter is $\lambda=0.95$.

The solutions $p(t)$ and $u(t)$ of the model depend on two control parameters: $\gamma$ representing the blowing pressure and $\zeta$ related to the opening of the embouchure at rest. In this work, the control parameter $\zeta$ is always constant and equal to 0.5. Using the variables $p^+=\frac{1}{2}\left(p+u\right)$ and $p^-=\frac{1}{2}\left(p-u\right)$ (outgoing and incoming pressure waves respectively) instead of the variables $p$ and $u$, the nonlinear characteristic of the exciter is written:

\begin{equation}
p^+=f\left(-p^-,\gamma\right).
\label{eq:NonLinRelpppm}
\end{equation}

Outgoing and incoming pressure waves are also related through the reflection function $r(t)$: 

\begin{equation}
 p^-(t) = \left(r * p^+\right)(t)=-\lambda p^+(t-\tau).
\label{eq:reflection_function}
\end{equation}

Finally, by combining equations (\ref{eq:NonLinRelpppm}) and (\ref{eq:reflection_function}) and using a discrete time formulation (the discretization is done at regular intervals $\tau$) and noting $p^+(n\tau)=p^+_n$ and $p^-(n\tau)=p^-_n$, we obtain the following iterated map~\cite{Maga1986,McIntyre83:JASA,MechOfMusInst}:

\begin{equation}
p^+_n=f\left(\lambda p^+_{n-1},\gamma \right)=G\left(p^+_{n-1},\gamma \right),
\label{e:static_map}
\end{equation}
with, by definition: $G(x)\equiv f(\lambda x)$. The function $G$ can be written explicitly for $\zeta<1$ (see Taillard \textit{et al.}~\cite{NonLin_Tail_2010}).

When the control parameter $\gamma$ is constant, for low values of $\gamma$ the solution of eq.~(\ref{e:static_map}) stabilizes at an equilibrium point which corresponds to the fixed point $p^{+*}(\gamma)$ of the iterated function $G$. For a critical value  $\gamma_{st}$, namely the static bifurcation point (also called the static oscillation threshold) a flip bifurcation~\cite{KuzAppBif2004} occurs, i.e.

\begin{equation}
G'\left(p^{+*}(\gamma_{st})\right) =-1,
\label{Jacoc2}
\end{equation}
leading to a 2-valued periodic regime that corresponds to sound production.

For the lossless model (i.e. $\lambda=1$) the static oscillation threshold is equal to $\gamma_{st}=1/3$. If $\lambda<1$, $\gamma_{st}$ is larger than $1/3$, an expression of $\gamma_{st}$ is given by Kergomard \textit{et al.}~\cite{KergoActa2000}.

\subsection{Dynamic bifurcation}

\begin{figure*}
\centering
\subfigure[lossless: $\lambda=1$]{\includegraphics[width=81mm,keepaspectratio=true]{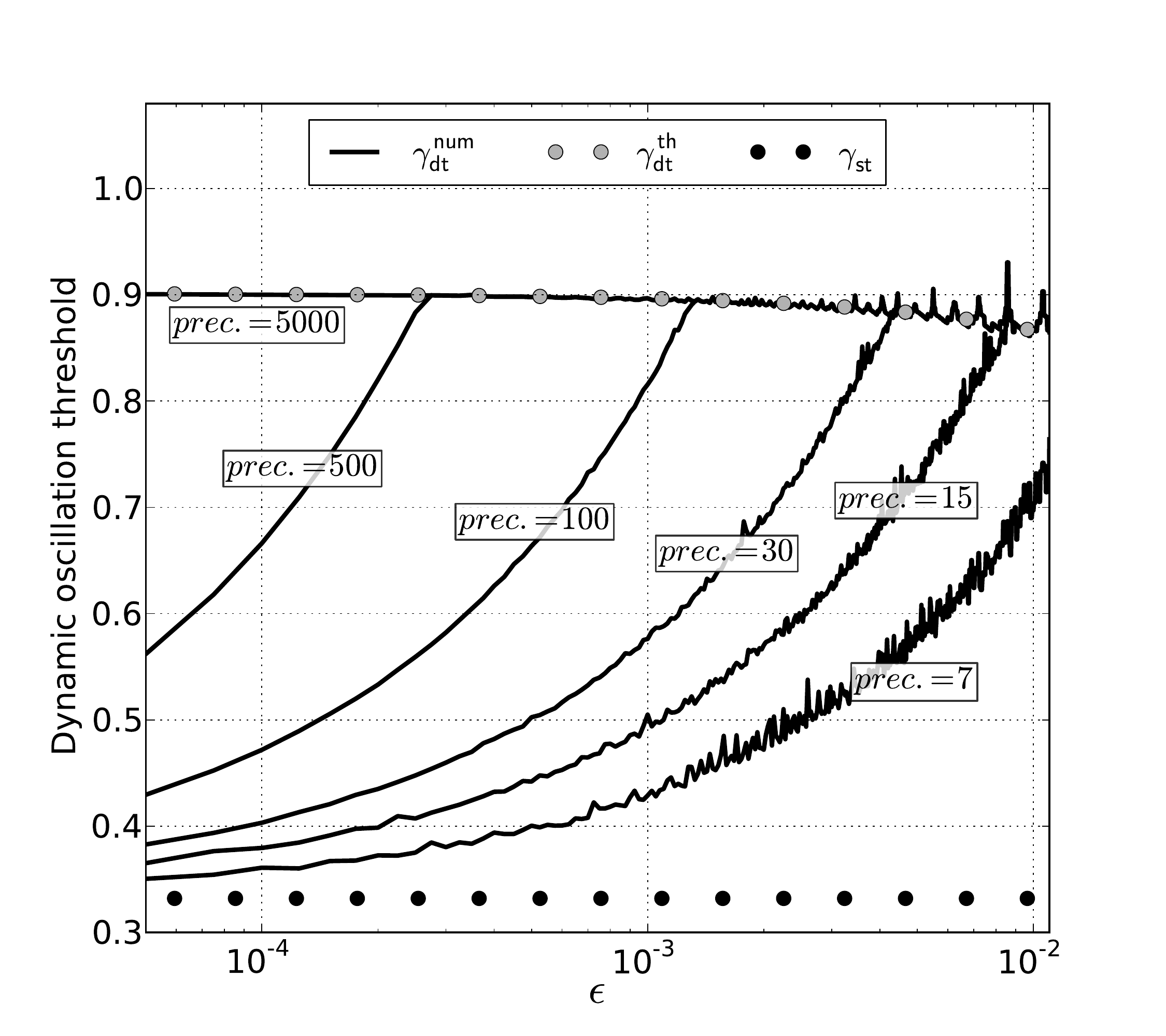}\label{fig:seuidynnoa}}
\subfigure[$\lambda=0.95$]{\includegraphics[width=81mm,keepaspectratio=true]{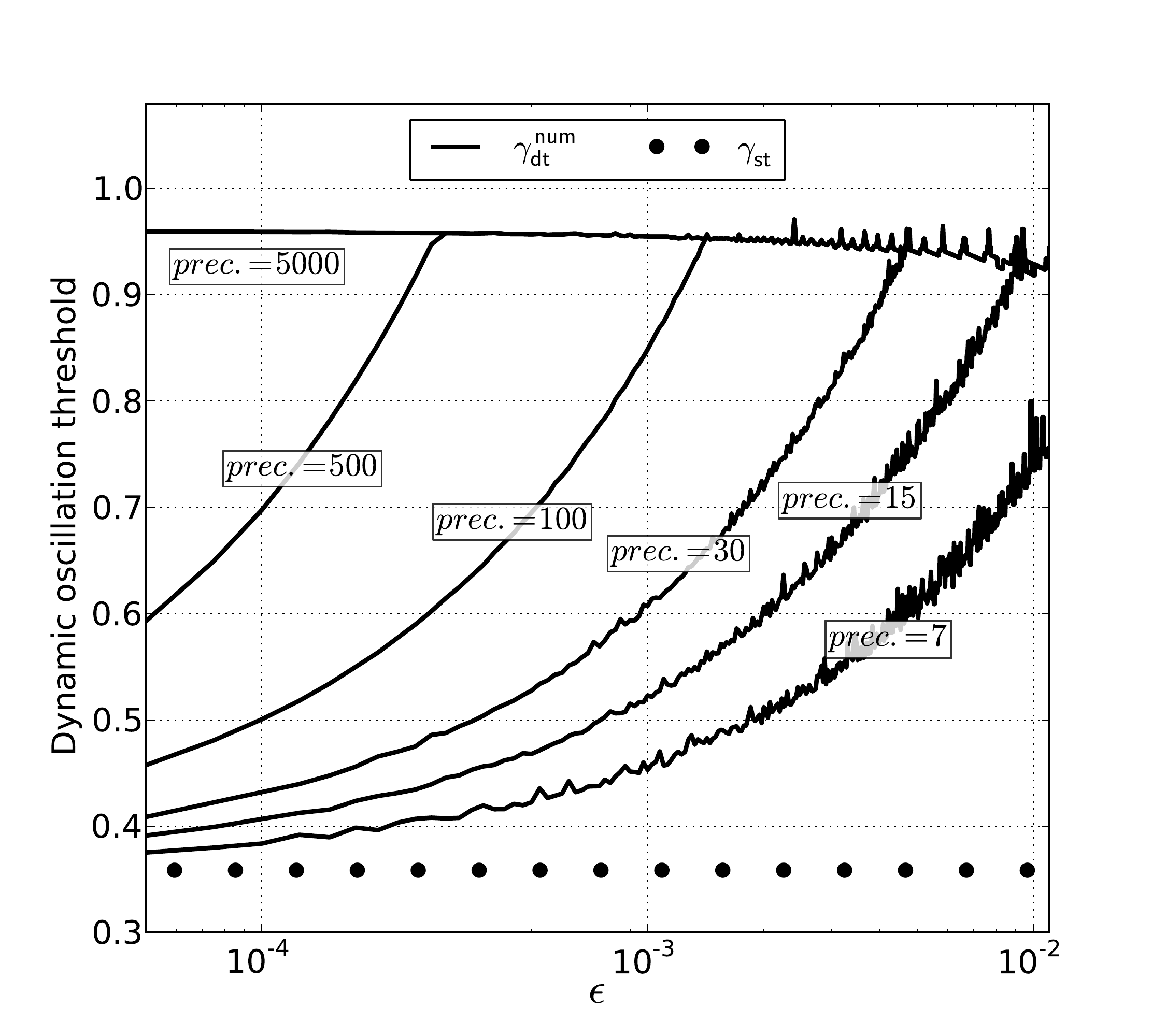}\label{fig:seuidynnob}}
\caption{Graphical representation of $\gamma_{dt}^{num}$ for different precisions (prec.~= 7, 15, 30, 100, 500 and 5000) with respect to the slope $\epsilon$ and for $\gamma_0 = 0$. Results are also compared to analytical \textit{static} and \textit{dynamic} thresholds: $\gamma_{st}$ and $\gamma_{dt}^{th}$. (a) lossless model: $\lambda=1$ and (b) typical losses in a cylindrical clarinet, $\lambda=0.95$.}
\label{fig:seuidynno}
\end{figure*}

For a linearly increasing control parameter $\gamma$,  eq.~(\ref{e:static_map}) is replaced by eq.~(\ref{dynsys_pp_a}) and (\ref{dynsys_pp_b}) :

\begin{subnumcases}{\label{dynsys_pp}}
p^+_n=G\left(p^+_{n-1},\gamma_n\right)\label{dynsys_pp_a}\\
\gamma_{n}=\gamma _{n-1} +\epsilon.\label{dynsys_pp_b}
\end{subnumcases}

The theory derived in section \ref{sec:SeDynNoTh} requires that the parameter $\gamma$ increases slowly, hence $\epsilon$ is considered arbitrarily small ($\epsilon\ll 1$).

Because of the time variation of the control parameter $\gamma$, the system~(\ref{dynsys_pp}) undergoes a bifurcation delay: the bifurcation point corresponding to the birth of the oscillations is shifted from the \textit{static oscillation threshold} $\gamma_{st}$~\cite{dalmont:3294} to the \textit{dynamic oscillation threshold} $\gamma_{dt}$~\cite{BergeotNLD2012}. The previous article by the authors~\cite{BergeotNLD2012} provides an analytical study of the dynamic flip bifurcation of the clarinet model (i.e.~system (\ref{dynsys_pp})) in the case where $\lambda=1$. The method is based on applications of dynamic bifurcation theory proposed by Baesens~\cite{Baesens1991}. The main focus of this work is on the properties of the dynamic oscillation threshold, recalled hereafter.

The trajectory of the system in the phase space (here constituted of a single variable $p^+$) through time is called the \emph{orbit}. The dynamic oscillation threshold is defined as the value of $\gamma$ for which the orbit escapes from the neighborhood of the \textit{invariant curve} $\phi(\gamma,\epsilon)$. This definition is different from the one used in~\cite{BergeotNLD2012} where the dynamic threshold was defined as the value of $\gamma$ for which the orbit starts to oscillate.

\begin{figure}[t!]
\centering
\subfigure[Numerical precision is fixed (prec. = 50). $\gamma_{dt}^{num}$ is computed for $\epsilon=10^{-4}$ and $3\cdot10^{-4}$.]{\includegraphics[width=77mm,keepaspectratio=true]{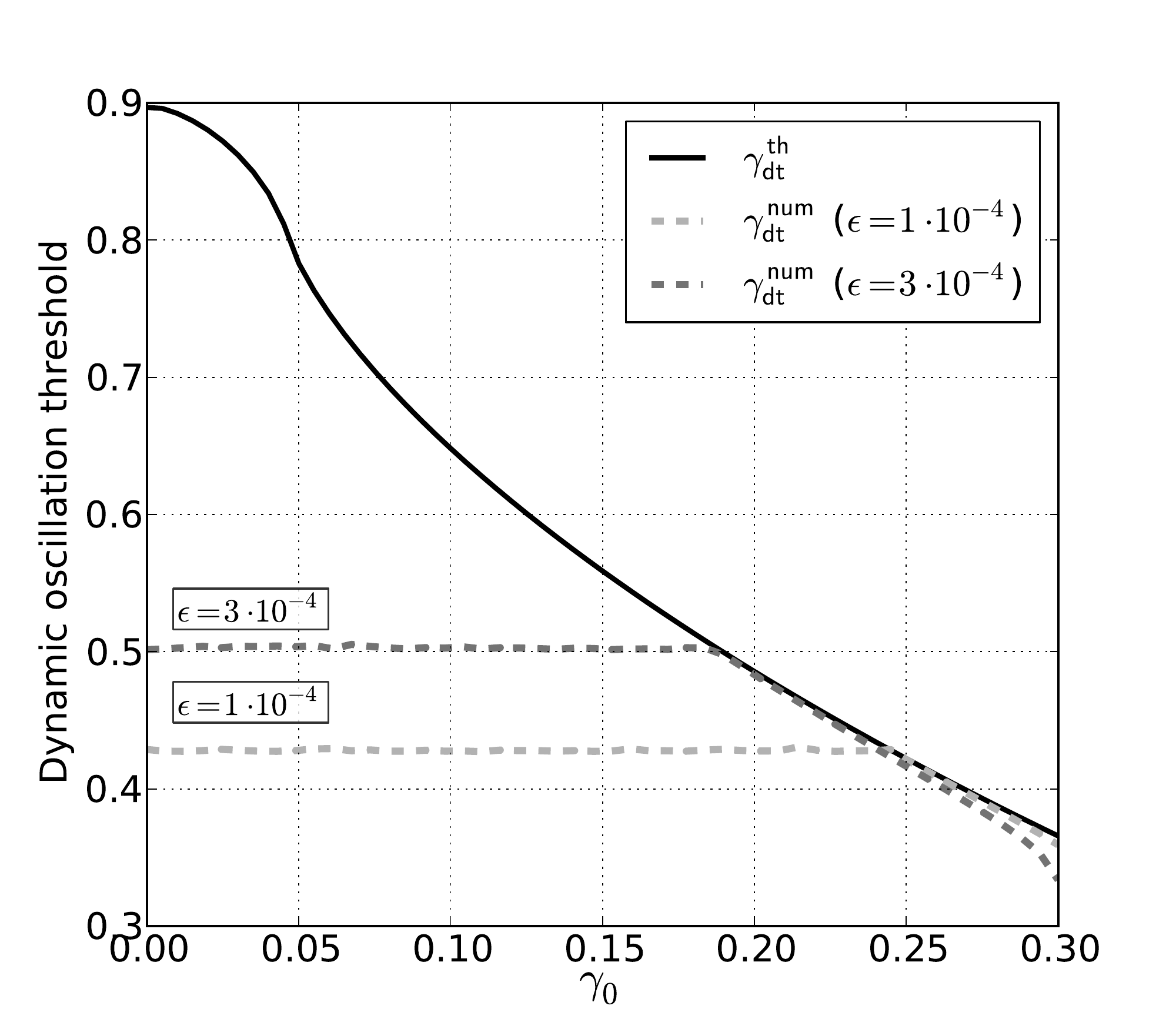}\label{fig:seuidynnoVScia}}
\subfigure[The increase rate of $\gamma$ is fixed ($\epsilon = 3\cdot 10^{-4}$). $\gamma_{dt}^{num}$ is computed for numerical precisions equal to 15 and 100.]{\includegraphics[width=77mm,keepaspectratio=true]{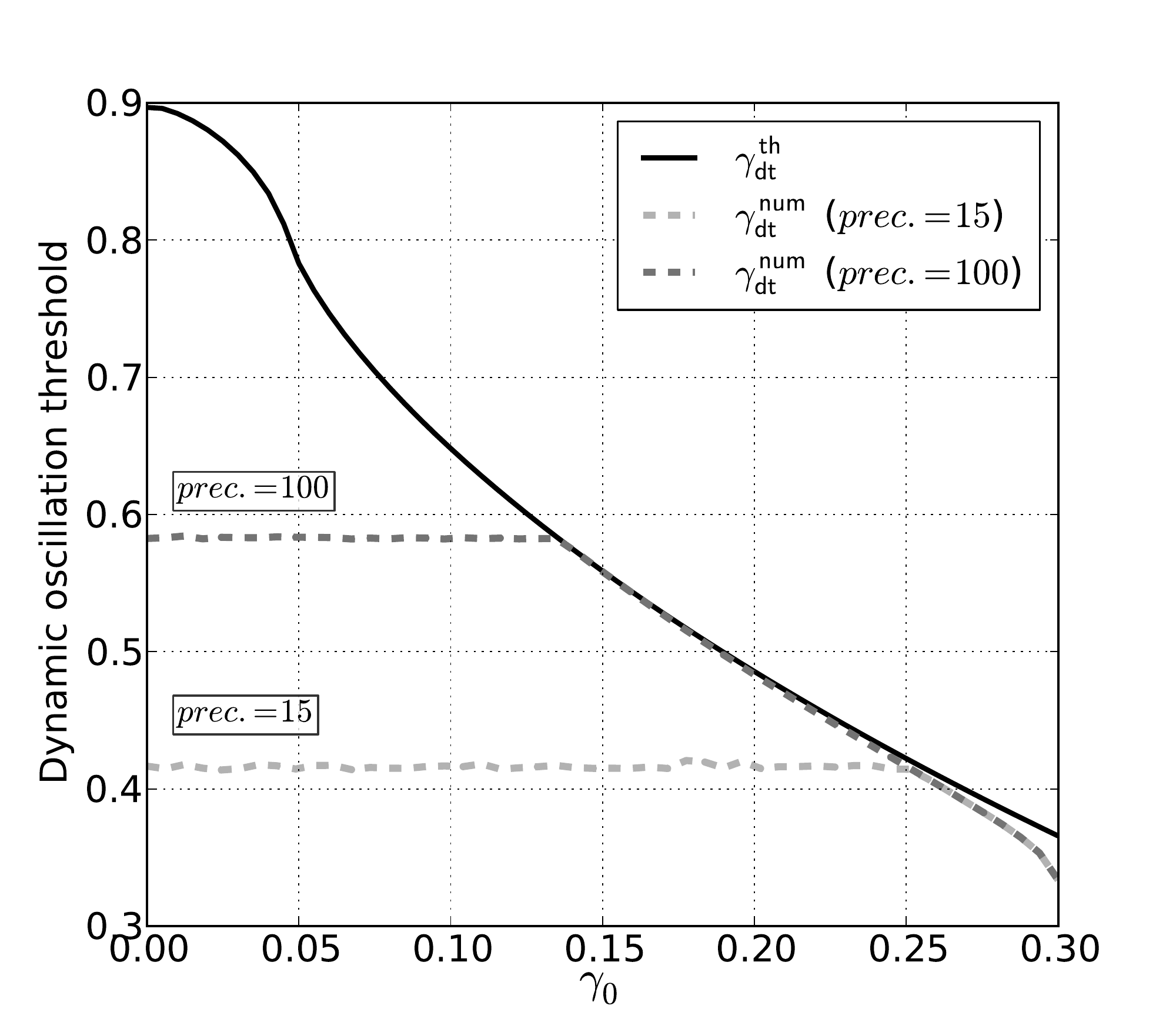}\label{fig:seuidynnoVScib}}
\caption{Plot of $\gamma_{dt}$ as a function of the initial condition $\gamma_0$. Solid lines are the theoretical prediction $\gamma_{dt}^{th}$ calculated from equation (\ref{dynoscthre_2}). Dashed line represent the values $\gamma_{dt}^{num}$.}
\label{fig:seuidynnoVSci}
\end{figure}

The invariant curve is the nonoscillating solution of the system (\ref{dynsys_pp}). It plays the role of an attractor for variable parameters similarly to the role of the fixed point in a static case. The invariant curve is written as a function of the parameter, invariant under the mapping (\ref{dynsys_pp}) and thus satisfying the following equation:
\begin{equation}
\phi(\gamma,\epsilon)=G\left(\phi(\gamma-\epsilon,\epsilon),\gamma\right).
\label{eqdiff_1}
\end{equation}

The procedure to obtain the theoretical estimation $\gamma_{dt}^{th}$ of the dynamic oscillation threshold is as follows: a theoretical expression of the invariant curve is found for a particular (small) value of the increase rate $\epsilon$ (i.e. $\epsilon \ll1$). The system~(\ref{dynsys_pp}) is then expanded into a first-order Taylor series around the invariant curve and the resulting linear system is solved analytically. Finally, $\gamma_{dt}^{th}$ is derived from the analytic expression of the orbit.

The dynamic oscillation threshold $\gamma_{dt}^{th}$ is defined by~\cite{BergeotNLD2012}:
\begin{equation}
\int_{\gamma_0+\epsilon}^{\gamma_{dt}^{th}+\epsilon}\ln\left| \partial_xG\left(\phi(\gamma'-\epsilon),\gamma'\right)\right|d\gamma'=0,
\label{dynoscthre_2}
\end{equation}
where $\gamma_0$ is the initial value of $\gamma$ (i.e.~the starting value of the linear ramp). Two important remarks can be made on this expression (Fig.~6 of ~\cite{BergeotNLD2012}):
\begin{itemize}
\item $\gamma_{dt}^{th}$ does not depend on the slope of the ramp $\epsilon$, provided that $\epsilon$ is small enough,
\item $\gamma_{dt}^{th}$  depends on the initial value $\gamma_0$ of the ramp.
\end{itemize}
These properties are also observed in numerical simulations with very high precision.

\subsection{Problem statement \label{ss:pbstate}}

The above theoretical predictions converge towards the observed simulation results for very high numerical precision (typically when thousands of digits are considered in the simulation). Figure\footnote{Figure~\ref{fig:seuidynnoa} is a plot similar to figure 10 of~\cite{BergeotNLD2012}. The only difference is that the bifurcation point $\gamma_{dt}^{num}$ estimated on the simulation results is here defined by the point where the orbit leaves the neighborhood of the invariant curve. The motivation for this choice will appear clearly in section \ref{s:pr_noise} where random variables are considered.} \ref{fig:seuidynnoa} shows that for the usual double-precision of CPUs (around 15 decimals), theoretical predictions of the dynamic bifurcation point $\gamma_{dt}^{th}$ are far from thresholds estimated on the numerical simulation results $\gamma_{dt}^{num}$. In particular, the numerical bifurcation point $\gamma_{dt}^{num}$ depends on the slope $\epsilon$, in contrast with the theoretical predictions $\gamma_{dt}^{th}$.

Moreover, figure~\ref{fig:seuidynnoVSci} reveals that for a low numerical precision (though even significantly higher than typical precisions used in numerical simulations), the dependence of  the bifurcation point on the initial value $\gamma_0$ is lost over a wide range of $\gamma_0$.

The minimum precision for which round-off errors do not affect the behavior of the system depends on the precision itself and on the relative magnitude of the slope $\epsilon$ and the initial condition $\gamma_0$.  Indeed, figure~\ref{fig:seuidynno} shows that, beyond a certain value of $\epsilon$ all curves join the one with highest precision. Curves for even higher precisions would overlap, allowing to conclude that they are representative of an infinitely precise case. As shown in figure~\ref{fig:seuidynnoVSci}, for given values of $\epsilon$ and of the numerical precision beyond a certain value of $\gamma_0$, the theoretical result $\gamma_{dt}^{th}$ allows to obtain a good prediction of the bifurcation delay.

As a conclusion, the theoretical results obtained in~\cite{BergeotNLD2012} are not able to predict the behavior of numerical simulations carried out at usual numerical precision. The aim of this paper is to show how the numerical precision can be included in a theoretical model that correctly describes numerical simulations. Firstly, it is shown that the model computed with a finite precision behaves similarly to the model with an ad-hoc additive white noise. This is done in the next section.  Then, using theoretical results given by Baesens~\cite{Baesens1991}, a modified expression describing the behavior of the model affected by noise (section \ref{sec:SeDynNoTh}) is proposed.

In figure~\ref{fig:seuidynnob}, system (\ref{dynsys_pp}) is simulated with $\lambda=0.95$, a typical value to take into account losses in the cylindrical clarinet considered in this paper. The effect of the losses is to increase the dynamic threshold, as for the static one.  However, the behavior of the lossless model and that of system with losses are qualitatively the same. Therefore, for sake of simplicity and without loss of generality, following analytical calculation and numerical simulations are performed using $\lambda=1$.


\section{Finite precision versus additive white noise \label{s:pr_noise}}

Differences between theoretical predictions and numerical simulations highlighted in the previous section are due to round-off errors that accumulate for finite precisions. The aim of this section is to test whether round-off errors of the computer can be modeled as an additive independent and identically distributed random variable (referred to as an additive white noise). This result is used in section \ref{sec:SeDynNoTh} to derive theoretical predictions of the dynamic bifurcation point $\gamma_{dt}$.

\begin{figure}[H]
\centering
\includegraphics[width=77mm,keepaspectratio=true]{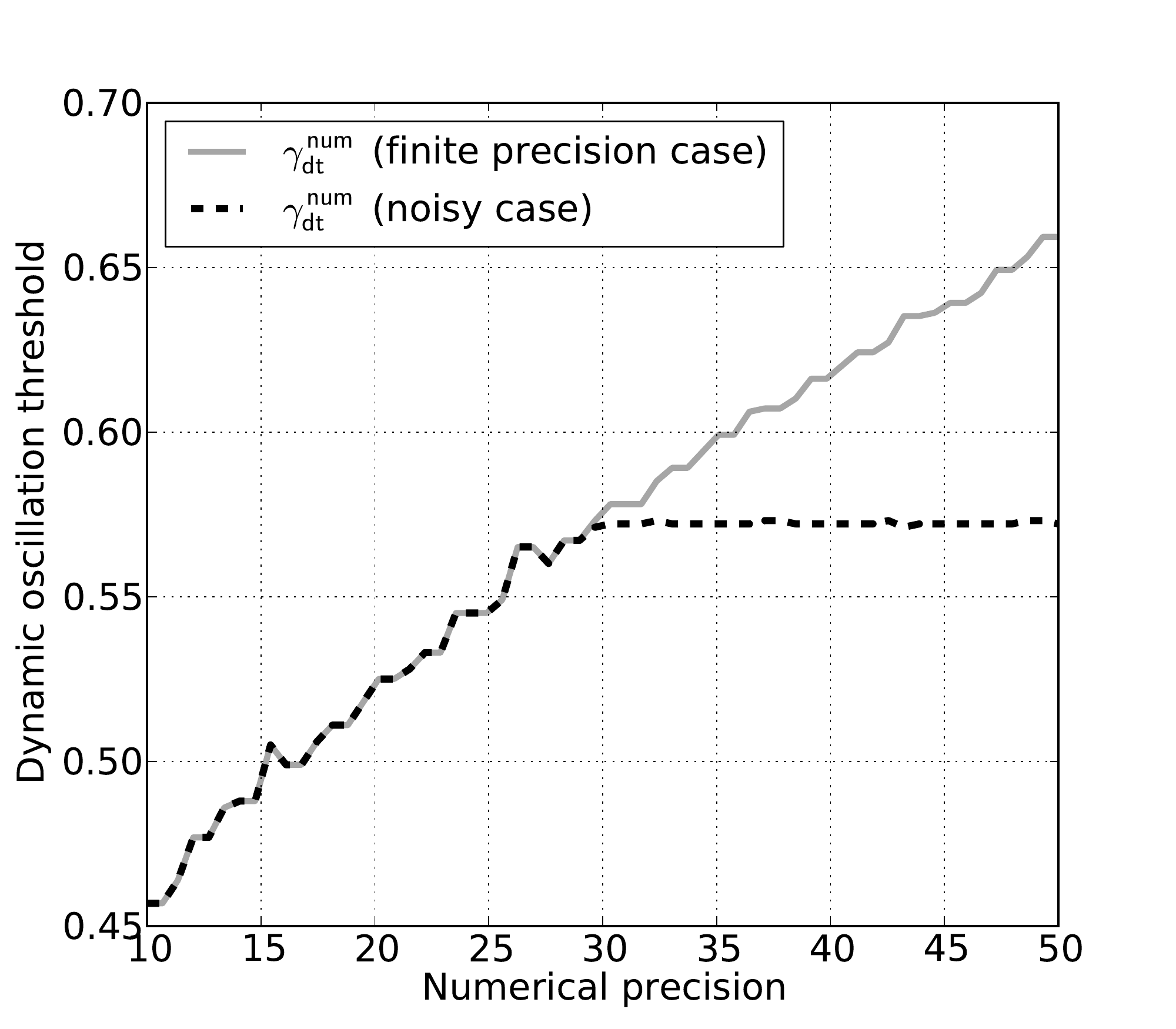}
\caption{Comparison of the dynamic threshold $\gamma_{dt}^{num}$ obtained in numerical simulations of a clarinet model in finite precision case (\ref{dynsys_pp}) and noisy case (\ref{dynsys_pp_no}) with a noise of level  $\sigma=10^{-30}$. The dynamic threshold of oscillation obtained over an average of 20 runs is plotted against the precision used in the simulations, showing that beyond a precision of about $\sigma$, the system affected with noise is insensitive to the precision.}
\label{fig:seuidyn:precVSrd0}
\end{figure}

\subsection{Results \label{s:pr_noise1}}

\begin{figure*}[t]
\centering
\includegraphics[width=115mm,keepaspectratio=true]{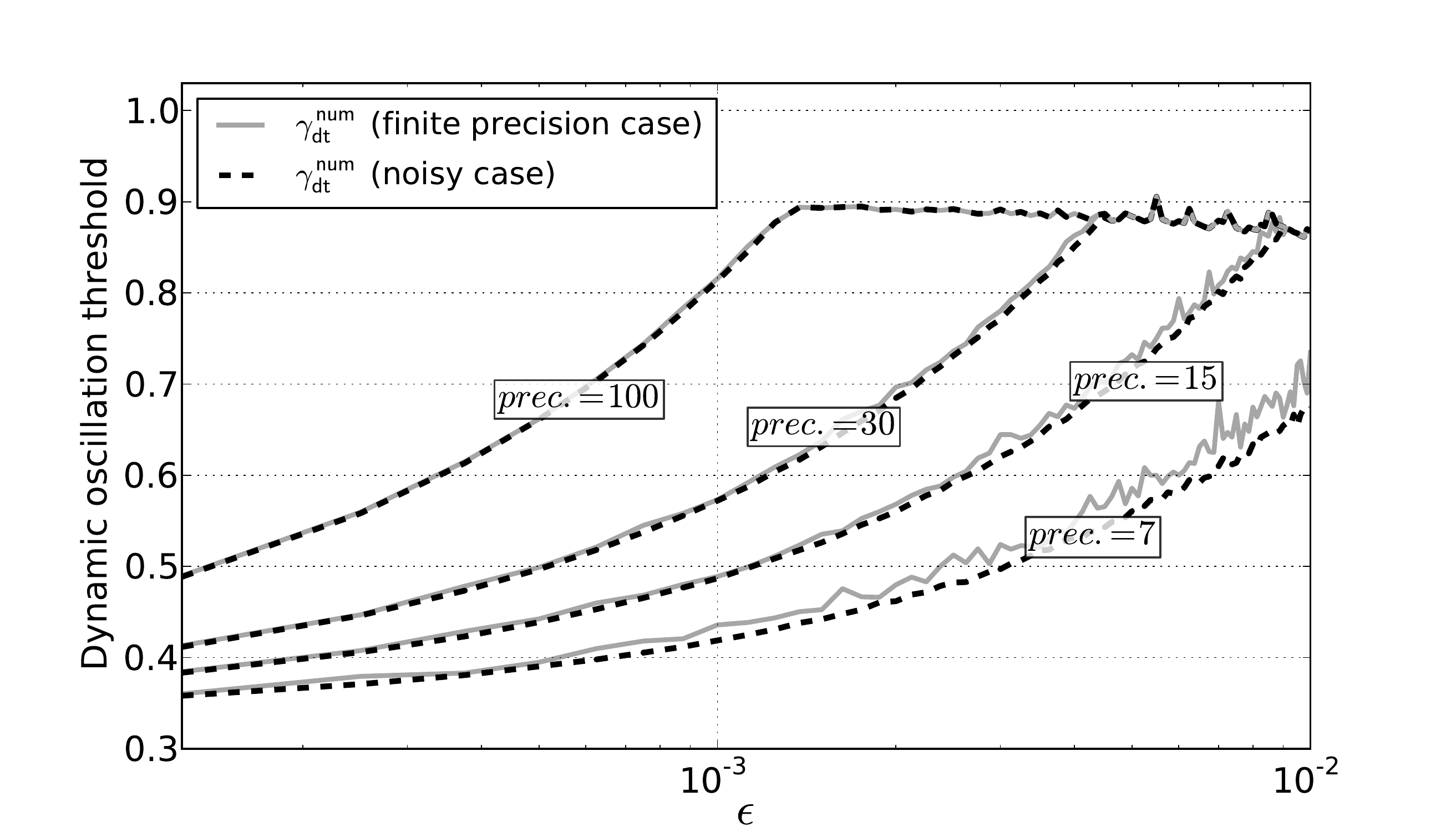}
\caption{Comparison between $\gamma_{dt}^{num}$ computed for \textit{finite precision cases} and for \textit{noisy cases}. For both cases and for each value of $\epsilon$ we compute the average of the signals $w_n = p^+_n-\phi(\gamma_n)$ obtained over 20 runs. Then, $\gamma_{dt}^{num}$ is calculated on the resulting signal. The numerical precisions used to simulate the finite precision cases are 7, 15, 30 and 100 decimal digits. $\gamma_0=0$.}
\label{fig:seuidyn:precVSrd}
\end{figure*}

Two numerical results are compared. The first is the simulation of the system (\ref{dynsys_pp}) using a numerical precision $pr_1$ (hereafter referred as a \textit{finite precision case}). The second one (hereafter referred as a \textit{noisy case}) is the simulation of the following stochastic system of difference equations:

\begin{subnumcases}{\label{dynsys_pp_no}}
p^+_n=G\left(p^+_{n-1},\gamma_n\right)+\xi_n \label{dynsys_pp_no_a}\\
\gamma_{n}=\gamma _{n-1} +\epsilon,
\end{subnumcases}
where $\xi_n$ is a uniformly distributed stochastic variable with an expected value equal to zero (i.e. $\EV{\xi_n}=0$) and a level $\sigma$ defined by:
\begin{equation}
\EV{\xi_m \xi_n} \, =\sigma^2\delta_{mn},
\label{eq:noiselev}
\end{equation}
where $\delta_{mn}$ is the Kronecker delta. The definition of the expected value $\mathbb{E}$ is provided in \cite{IntroProbShel}. For comparison with the finite precision case the noise level $\sigma$ is equal to $10^{-pr_1}$.

The bifurcation point $\gamma_{dt}^{num}$ estimated on the simulations is defined as the value of $\gamma$ for which the orbit leaves the neighborhood of the invariant curve. Since the mean value of the white noise $\xi_n$ is zero, the relevant quantity to study is the mean square deviation of the orbit from the invariant curve. Therefore, $\gamma_{dt}^{num}$ is reached when:

\begin{equation}
\sqrt{\EV{w_n^2}} = \epsilon,
\label{eq:crit2}
\end{equation}
where $w_n = p^+_n-\phi(\gamma_n,\epsilon)$ describes the distance between the actual orbit and the invariant curve. Among other possible criteria, the condition (\ref{eq:crit2}) is chosen because it is also used in the analytical calculation made in section \ref{sec:SeDynNoTh}.

To simplify the notation, in the rest of the document the invariant curve will be noted $\phi(\gamma)$. Its dependency on parameter $\epsilon$ is no longer explicitly stated.

In figures \ref{fig:seuidyn:precVSrd0} and \ref{fig:seuidyn:precVSrd}, $\gamma_{dt}^{num}$ is estimated in the finite precision case and in the noisy case. In both cases an average is made on $w_n^2$ obtained in 20 different simulations. Then, $\gamma_{dt}^{num}$ is calculated on the mean signal using equation (\ref{eq:crit2}).

In figure \ref{fig:seuidyn:precVSrd0}, $\gamma_{dt}^{num}$ is plotted with respect to the numerical precision for which both systems (\ref{dynsys_pp}) and (\ref{dynsys_pp_no}) are simulated. The noise level $\sigma$ of the noisy case modeled by the system~(\ref{dynsys_pp_no}) is equal to $10^{-30}$. For numerical precision below $-\log_{10}(\sigma)=30$, the noise level is smaller than round-off errors of the computer. In these situations, the effect of the additive noise in system~(\ref{dynsys_pp_no}) is hidden by the effect of the round-off errors of the computer. The consequence is that the thresholds computed in finite precision case and in noisy case are equals. For numerical precisions higher than 30, $\gamma_{dt}^{num}$ computed on system (\ref{dynsys_pp_no}) is constant because the influence of the round-off errors is now hidden by the additive noise which have a fixed level. Figure \ref{fig:seuidyn:precVSrd0} shows that the transition between the regime for which the round-off error effect prevails over the additive noise  affect and the regime for which the opposite occurs is abrupt. Therefore, the region where mixed effects of both round-off errors and additive noise play a role is very narrow. However, to avoid any influence of the numerical precision, the system (\ref{dynsys_pp_no}) is simulated using a precision $pr_2 = 2pr_1$.

Figure  \ref{fig:seuidyn:precVSrd} confirms that the kind of noise introduced in the stochastic system can correctly describe the influence of a finite precision. Indeed, with the exception of the smallest precision ($pr_1=7$), the curves are nearly superimposed. Hence, in the next section, the stochastic system (\ref{dynsys_pp_no}) is studied theoretically in order to predict results of numerical simulations of system (\ref{dynsys_pp}) with finite precision.

\subsection{Relevance of numerical results}

\begin{table*}[t]
\centering
{\small
\rowcolors{3}{}{gray!25}
\caption{Mean value $\EV{w_n^2}$ and  standard deviation $\STD{w_n^2}$ of the signal $w_n^2$, calculated at the dynamic threshold $\gamma_{dt}^{num}$. Mean values $\sqrt{\EV{w_n^2}}$ and standard deviation $\STD{\gamma_{dt}^{num,i}}$ of dynamic thresholds $\gamma_{dt}^{num,i}$ ($i\in [1,20]$) calculated on each run. All results are calculated for $\sigma=10^{-7}$ and $10^{-15}$ and for $\epsilon=10^{-4}$, $10^{-3}$ and $10^{-3}$.}
\label{tab:QualNumRes}       
\renewcommand{\arraystretch}{1.6}
\begin{tabular}{ccc|c|c|c|c|c|c|}
\cline{4-9}
& & & \multicolumn{3}{ c| }{\multirow{2}{*}{$\sigma=10^{-7}$}} &  \multicolumn{3}{ c| }{\multirow{2}{*}{$\sigma=10^{-15}$}} \\
& & & \multicolumn{3}{ c| }{} &  \multicolumn{3}{ c| }{}\\ \hline 
\multicolumn{3}{ |p{4cm}| }{$\epsilon$} & $10^{-4}$ & $10^{-3}$ & $10^{-2}$ & $10^{-4}$ & $10^{-3}$ & $10^{-2}$ \\ \hline
\multicolumn{3}{ |p{4cm}| }{$\EV{w_n^2}$ at $\gamma_{dt}^{num}$} & $1.01\cdot10^{-8}$ & $1.18\cdot10^{-6}$ & $2.83\cdot10^{-4}$ & $1.20\cdot10^{-8}$ & $1.46\cdot10^{-6}$ & $5.46\cdot10^{-4}$ \\ \hline
\multicolumn{3}{ |p{4cm}| }{$\text{STD}[w_n^2]$ at $\gamma_{dt}^{num}$} & $1.24\cdot10^{-8}$ & $1.56\cdot10^{-6}$ & $3.02\cdot10^{-4}$ & $2.07\cdot10^{-8}$ & $1.86\cdot10^{-6}$ & $1.25\cdot10^{-4}$ \\ \hline
\multicolumn{3}{ |p{4cm}| }{$\gamma_{dt}^{num}$ estimated on $\sqrt{\EV{w_n^2}}$} & $0.354$ & $0.418$ & $0.673$ & $0.377$ & $0.488$ & $0.857$ \\ \hline
\multicolumn{3}{ |p{4cm}| }{$\EV{\gamma_{dt}^{num,i}}$} & $0.355$ & $0.421$ & $0.677$ & $0.378$ & $0.490$ & $0.856$ \\ \hline
\multicolumn{3}{ |p{4cm}| }{$\STD{\gamma_{dt}^{num,i}}$} & $0.002$ & $0.005$ & $0.014$ & $0.001$ & $0.003$ & $0.005$ \\ \hline
\end{tabular}
}
\end{table*}

To investigate the relevance of the numerical results, several indicators are calculated. First, the standard deviation $\STD{w_n^2}$ of the signal $w_n^2$ is calculated at  the dynamic threshold $\gamma_{dt}^{num}$ and compared to $\EV{w_n^2}$, also calculated at $\gamma_{dt}^{num}$. Secondly, the dynamic threshold is calculated on each run. We obtain 20 values, noted $\gamma_{dt}^{num,i}$ ($i\in [1,20]$). The mean value $\EV{\gamma_{dt}^{num,i}}$ is compared to the value $\gamma_{dt}^{num}$, estimated on the mean signal $\sqrt{\EV{w_n^2}}$ (see section~\ref{s:pr_noise1} where this numerical estimation method of $\gamma_{dt}^{num}$ is used because it also used in analytical calculations in section~\ref{sec:SeDynNoTh}.). The standard deviation $\STD{\gamma_{dt}^{num,i}}$ is calculated too.

Results are presented in table~\ref{tab:QualNumRes}. The mean value $\EV{w_n^2}$ and the standard deviation $\STD{w_n^2}$ at the dynamic threshold have the same order of magnitude. This suggests a bad repeatability of the numerical simulations. However, at the dynamic threshold, $w_n$ diverges sharply and a large deviation of it does not necessarily imply a large deviation of the dynamic threshold. The standard deviation $\STD{\gamma_{dt}^{num,i}}$, in table~\ref{tab:QualNumRes}, shows precisely a good repeatability of $\gamma_{dt}^{num,i}$.

\section{Analytical study  of the noisy dynamic case}
\label{sec:SeDynNoTh}

\subsection{General solution of the stochastic clarinet model}

This section introduces a formal solution of the stochastic model that is valid when the orbit is close to the invariant curve. Function $G$ in equation (\ref{dynsys_pp_no_a}) is expanded into a first-order Taylor series around the invariant curve. Using the variable $w_n = p^+_n-\phi(\gamma_n)$, the system (\ref{dynsys_pp_no}) becomes:

\begin{subnumcases}{\label{dynsys_pplin_no}}
w_n=w_{n-1} \partial_xG\left(\phi(\gamma_n-\epsilon),\gamma_n\right)+\xi_n \label{dynsys_pplin_no_a} \\
\gamma_{n}=\gamma _{n-1} +\epsilon.
\end{subnumcases}

The solution of equation (\ref{dynsys_pplin_no_a}) is \cite{bender78}:

\begin{multline}
w_n=w_0\prod_{k=1}^{n}\partial_xG\left(\phi(\gamma_k-\epsilon),\gamma_k\right)\\+\sum_{k=1}^{n}\xi_k\prod_{m=k+1}^{n} \partial_xG\left(\phi(\gamma_m-\epsilon),\gamma_m\right),
\label{exact_solution}
\end{multline}
where $w_0$ is the initial value of $w_n$.

Because the additive white noise $\xi_n$ has a zero-value mean, as in section \ref{s:pr_noise}, the relevant indicator is the mean square deviation of the orbit from the invariant curve: $\sqrt{\EV{w_n^2}}$. Equation (\ref{exact_solution}) squared becomes:

\begin{multline}\label{eq:squarew}
w_n^2 = \left(w_0\prod_{k=1}^{n}\partial_xG\left(\phi(\gamma_k-\epsilon),\gamma_k\right)\right)^2\\
+ \left(\sum_{k=1}^{n}\xi_k\prod_{m=k+1}^{n} \partial_xG\left(\phi(\gamma_m-\epsilon),\gamma_m\right)\right)^2\\
+ 2w_0\sum_{k=1}^{n}\left(\prod_{j=1}^{n}\partial_xG\left(\phi(\gamma_j-\epsilon),\gamma_j\right)\right)\xi_k\\
\times \prod_{m=k+1}^{n} \partial_xG\left(\phi(\gamma_m-\epsilon),\gamma_m\right).
\end{multline}

Averaging has no effect on the first term of the right-hand side of equation (\ref{eq:squarew})  because the stochastic variable $\xi_n$ is not present. Using eq.~(\ref{eq:noiselev}), the average of the second term is simplified to:

\begin{equation}
\sigma^2\sum_{k=1}^{n}\left(\prod_{m=k+1}^{n} \partial_xG\left(\phi(\gamma_m-\epsilon),\gamma_m\right)\right)^2.
\end{equation}

Because $\EV{\xi_n}=0$, the average of the third term of the right-hand side of equation (\ref{eq:squarew}) is also equal to zero. Using the fact that a product can be expressed as an exponential of a sum of logarithms, the final expression of $\EV{w_n^2}$ is given by:

\begin{multline}
\EV{w_n^2} = \underbrace{w_0^2\left(\exp\left(\sum_{k=1}^{n}\ln\left| \partial_xG\left(\phi(\gamma_k-\epsilon),\gamma_k\right)\right|\right)\right)^2}_{A_n}\\+\underbrace{\sigma^2\sum_{k=1}^{n}\left(\exp\left[\sum_{m=k+1}^{n} \ln\left|\partial_xG\left(\phi(\gamma_m-\epsilon),\gamma_m\right)\right|\right]\right)^2}_{B_n}.
\label{exact_solution_measq_sim}
\end{multline}

The two terms of the right-hand side of equation (\ref{exact_solution_measq_sim}) are denoted $A_n$ and $B_n$.

Finally, using Euler's approximation, sums are replaced by integrals and the expressions of $A_n$ and $B_n$ become:

\begin{equation}
A_n \approx w_0^2\exp\left(\int_{\gamma_0+\epsilon}^{\gamma_n+\epsilon}2\ln\left| \partial_xG\left(\phi(\gamma'-\epsilon),\gamma'\right)\right|\frac{d\gamma'}{\epsilon}\right),
\label{An_cont}
\end{equation}

\begin{multline}
B_n \approx \frac{\sigma^2}{\epsilon} \int_{\gamma_0+\epsilon}^{\gamma_n+\epsilon} \\ \left\{ \exp\left(\int_{\gamma+\epsilon}^{\gamma_n+\epsilon}2\ln\left| \partial_xG\left(\phi(\gamma'-\epsilon),\gamma'\right)\right|\frac{d\gamma'}{\epsilon}\right)\right\}d \gamma.
\label{Bn_cont}
\end{multline}

$A_n$ corresponds to the precise case studied in~\cite{BergeotNLD2012} which leads to the theoretical estimation $\gamma_{dt}^{th}$ of the dynamic oscillation threshold for the system without noise (cf.~equation (\ref{dynoscthre_2})). $B_n$ is the contribution due to the noise.

The transform from discrete sums to integral can be questioned. Indeed, to transform the term $A_n$ in equation~(\ref{exact_solution_measq_sim}) to its integral form~(\ref{An_cont}), we assume that:
\begin{multline}
\sum_{k=1}^{n}\ln\left| \partial_xG\left(\phi(\gamma_k-\epsilon),\gamma_k\right)\right| \approx \\
\int_{\gamma_0+\epsilon}^{\gamma_n+\epsilon}\ln\left| \partial_xG\left(\phi(\gamma'-\epsilon),\gamma'\right)\right|\frac{d\gamma'}{\epsilon},
\label{equalitySutoIN}
\end{multline}
but if $\partial_xG\left(\phi(\gamma_i-\epsilon),\gamma_i\right)$ crosses over zero, one term in the sum comes close to $\ln(0) =-\infty$ and the equality~(\ref{equalitySutoIN}) is respected only for small enough values of $\epsilon$. Otherwise, in equation~(\ref{exact_solution_measq_sim}), we have: $A_n \rightarrow 0$ and $B_n \rightarrow \sigma^2$, and consequently $\sqrt{\EV{w_n^2}} \rightarrow \sigma$. In this case, for small noise levels, the difference between the orbit and the invariant curve comes close to zero, and as a result, the orbit needs more time to escape from the invariant curve neighborhood, i.e. the bifurcation delay is lengthened. This phenomenon is mentioned by Baesens~\cite{Baesens1991,Baesens1995} and Fruchard~\cite{Fruchard2007}. It can be observed for example in figure~\ref{fig:seuidynno} where peaks (i.e. larger bifurcation delay) are there on curves in the above right part of the figure. In the rest of the paper, we use integral form, because it allows analytical integrations of noise contribution $B_n$ which will be considered in the remaining of this section.

A first glance on equations (\ref{An_cont}) and (\ref{Bn_cont}) allows to explain observation made in Section \ref{ss:pbstate}. Indeed, comparing the expressions of $A_n$ and $B_n$, it possible to distinguish \cite{Baesens1991,Baesens1991Noise} two operating regimes, which, for a given value of $w_0$, depends on $\epsilon$, $\sigma$ and $\gamma_0$:

\begin{itemize}
\item $A_n \gg B_n$ \textbf{(deterministic regime)}

In this case the noise does not affect the bifurcation delay and the dynamic oscillation threshold can be determined by eq.~(\ref{dynoscthre_2}).
 
\item $A_n \ll B_n$  \textbf{(sweep-dominant regime)}

In this case, the bifurcation delay is affected by the noise. This regime is studied in the following section.
\end{itemize}

In Section \ref{ss:pbstate},  figures~\ref{fig:seuidynno} and \ref{fig:seuidynnoVSci}  
represent two different cases distinguished by the parameter values: in certain areas of the figures, the dynamic bifurcation threshold does not depend on $\epsilon$ but depends  on  $\gamma_0$, while in other areas the dynamic bifurcation threshold  depends on $\epsilon$ but is not dependent  on  $\gamma_0$. This observation may be interpreted as the existence of the two regimes detailed above: a \textbf{sweep-dominant regime} and  a \textbf{deterministic regime}. The transition between the two regimes occurs abruptly as observed  in figures~\ref{fig:seuidynno} and \ref{fig:seuidynnoVSci}.

\subsection{Theoretical expression of the dynamic oscillation threshold of the stochastic model}
\label{sec:42}

\begin{figure}[t]
\centering
\subfigure[$\zeta=0.2$]{\includegraphics[width=65mm,keepaspectratio=true]{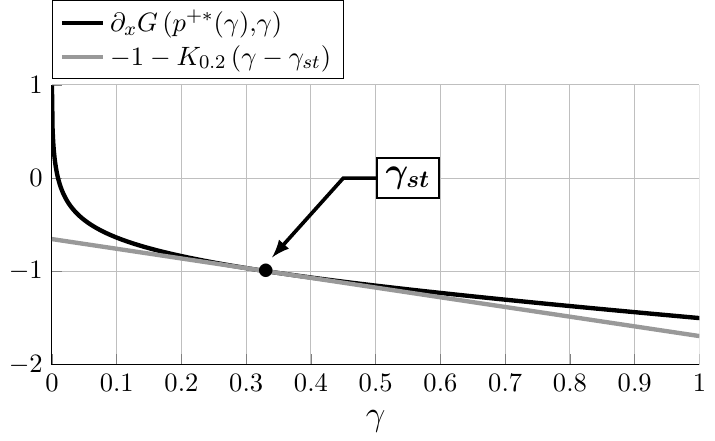}\label{fig:ggp_PFVSlina}}
\subfigure[$\zeta=0.5$]{\includegraphics[width=65mm,keepaspectratio=true]{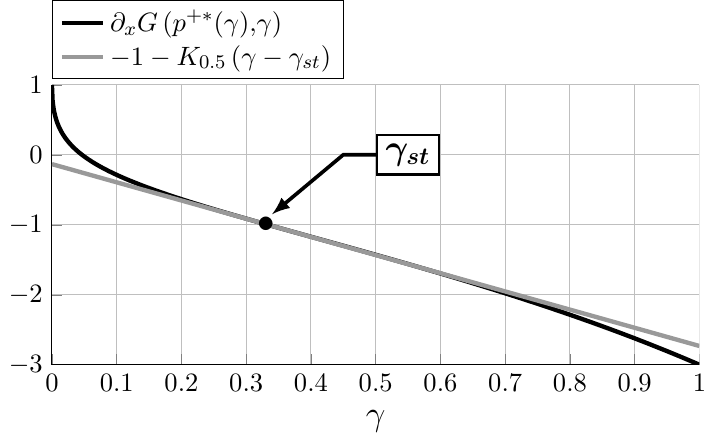}\label{fig:ggp_PFVSlinb}}
\subfigure[$\zeta=0.8$]{\includegraphics[width=65mm,keepaspectratio=true]{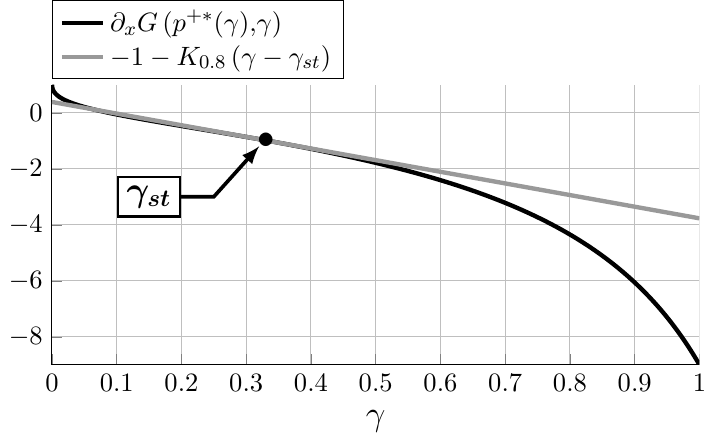}\label{fig:ggp_PFVSlinc}}
\caption{Graphical representation of $\partial_x G\left(p^{+*}(\gamma){,}\gamma \right)$ and its tangent function $-1-K\left(\gamma-\gamma_{st}\right)$ around the static oscillation threshold for $\zeta=0.2$, 0.5 and 0.8.}
\label{fig:ggp_PFVSlin}
\end{figure}

The next step is to find an approximate expression of the standard deviation $\sqrt{\EV{w_n^2}}$ for the sweep-dominant regime. In this regime, the term $A_n$ is negligible with respect to the contribution $B_n$ due to the noise, i.e. $\sqrt{\EV{w_n^2}}\approx\sqrt{B_n}$. The purpose is to obtain a statistical prediction of the dynamic oscillation threshold for the stochastic system, hereafter referred as $\hat{\gamma}_{dt}^{th}$. 

It is assumed that $\epsilon \ll 1$, which implies that the invariant curve $\phi(\gamma)$ and the curve $p^{+*}(\gamma)$ of the fixed points in eq.~(\ref{e:static_map}) are close \cite{BergeotNLD2012}, and allows the approximation:

\begin{equation}
\partial_xG\left(\phi(\gamma-\epsilon),\gamma\right) \approx \partial_xG\left(p^{+*}(\gamma),\gamma\right).
\label{H1}
\end{equation}

Moreover,  because of the noise, the bifurcation delay is expected to occur earlier, so that the dynamic oscillation threshold $\gamma_{dt}$ is assumed to be close\footnote{This hypothesis could be questioned because according to figures~\ref{fig:seuidynno} and \ref{fig:seuidynnoVSci}, even in the presence of noise, the bifurcation delay can be large. However, this hypothesis is required to carry out following calculations.
} to the static oscillation threshold $\gamma_{st}$. The term $\partial_xG\left(p^{+*}(\gamma),\gamma\right)$ is then expanded in a first-order Taylor series around the static oscillation threshold $\gamma_{st}$:

\begin{multline}
\partial_xG\left(p^{+*}(\gamma),\gamma\right) \approx  \underbrace{\partial_xG\left(p^{+*}(\gamma_{st}),\gamma_{st}\right)}_{\triangleq -1\,\text{: flip bifurcation}}\\
+\left(\gamma-\gamma_{st}\right) \underbrace{\partial_{xy}G\left(p^{+*}(\gamma_{st}),\gamma_{st}\right)}_{\text{noted}\, -K},
\label{TaExpGamSt}
\end{multline}
finally we have:

\begin{equation}
\partial_xG\left(p^{+*}(\gamma),\gamma\right) \approx  -1-K\left(\gamma-\gamma_{st}\right),
\label{dGpf_lin}
\end{equation}
which is used in equation (\ref{Bn_cont}). Figure \ref{fig:ggp_PFVSlin} shows the comparison between $\partial_x G\left(p^{+*}(\gamma){,}\gamma \right)$ and its tangent function $-1-K\left(\gamma-\gamma_{st}\right)$ around the static oscillation threshold for $\zeta=0.2$, 0.5 and 0.8. The linearisation appears as a good approximation of the function in a wide domain of $\gamma$ around the static oscillation threshold $\gamma_{st}$. For large values of the control parameter $\zeta$ (cf.~fig.~\ref{fig:ggp_PFVSlinc}) the linear approximation is valid over a narrower range of $\gamma$.

Using expression (\ref{dGpf_lin}) the integral
\begin{equation}
I_1=\int_{\gamma+\epsilon}^{\gamma_n+\epsilon}2\ln\left| \partial_xG\left(\phi(\gamma'-\epsilon),\gamma'\right)\right|\frac{d\gamma'}{\epsilon},
\end{equation}
contained in the expression (\ref{Bn_cont}) of $B_n$ becomes:

\begin{equation}
I_1=\frac{2K}{\epsilon}\int_{\gamma+\epsilon}^{\gamma_n+\epsilon} \left(\gamma'-\gamma_{st}\right) d\gamma'=
\frac{K}{\epsilon}\left[\left(\gamma'-\gamma_{st}\right)^2\right]_{\gamma+\epsilon}^{\gamma_n+\epsilon}.
\end{equation}

The small correction $\epsilon$ in the domain of integration can be neglected since $\epsilon \ll 1$. Therefore, we obtain:

\begin{multline}
I_1=\frac{K}{\epsilon}\left[\left(\gamma'-\gamma_{st}\right)^2\right]_{\gamma}^{\gamma_n}\\=
\frac{K}{\epsilon}\left[\left(\gamma_n-\gamma_{st}\right)^2-\left(\gamma-\gamma_{st}\right)^2\right].
\label{I1fin}
\end{multline}

\begin{figure}[H]
\centering
\subfigure[]{\includegraphics[width=83mm,keepaspectratio=true]{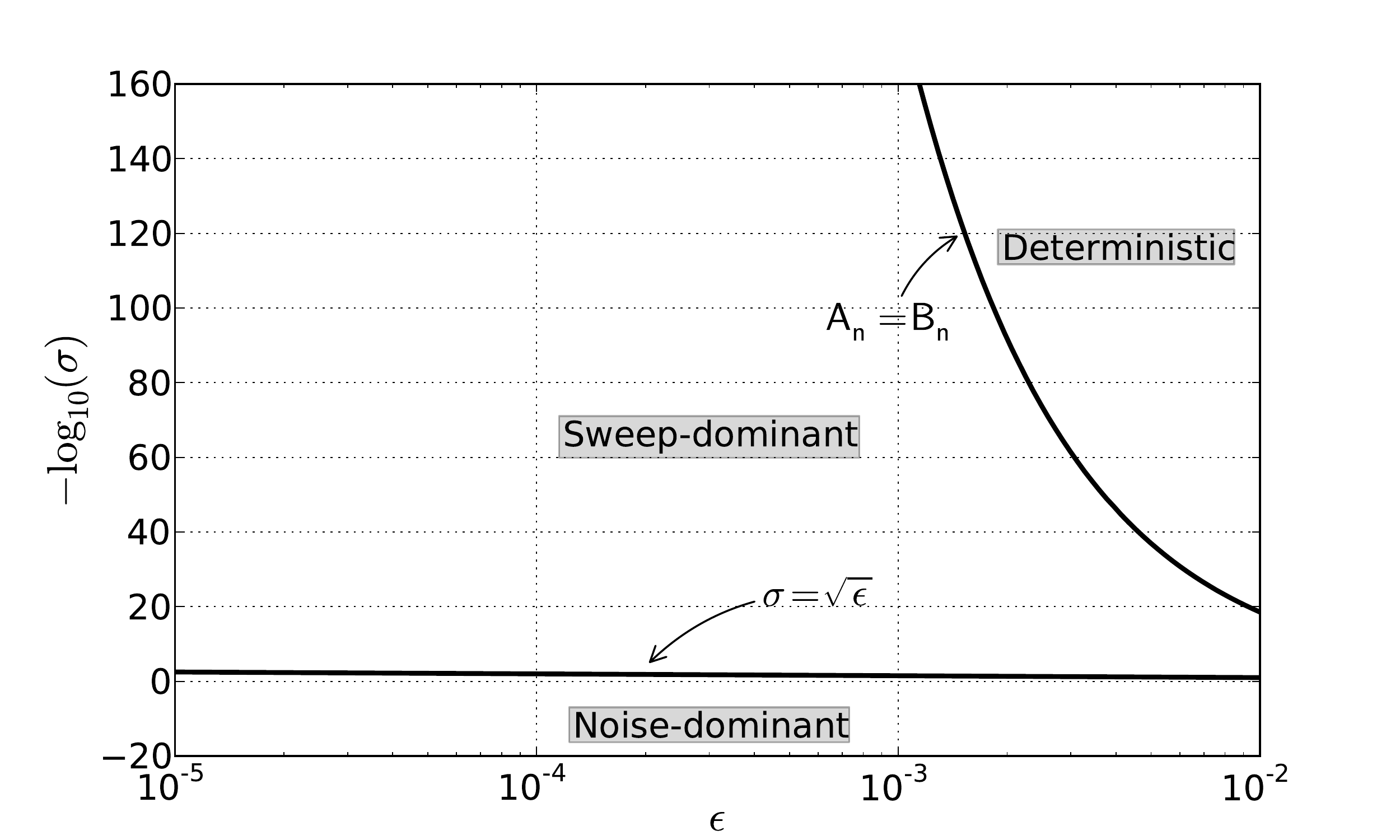}}
\subfigure[]{\includegraphics[width=83mm,keepaspectratio=true]{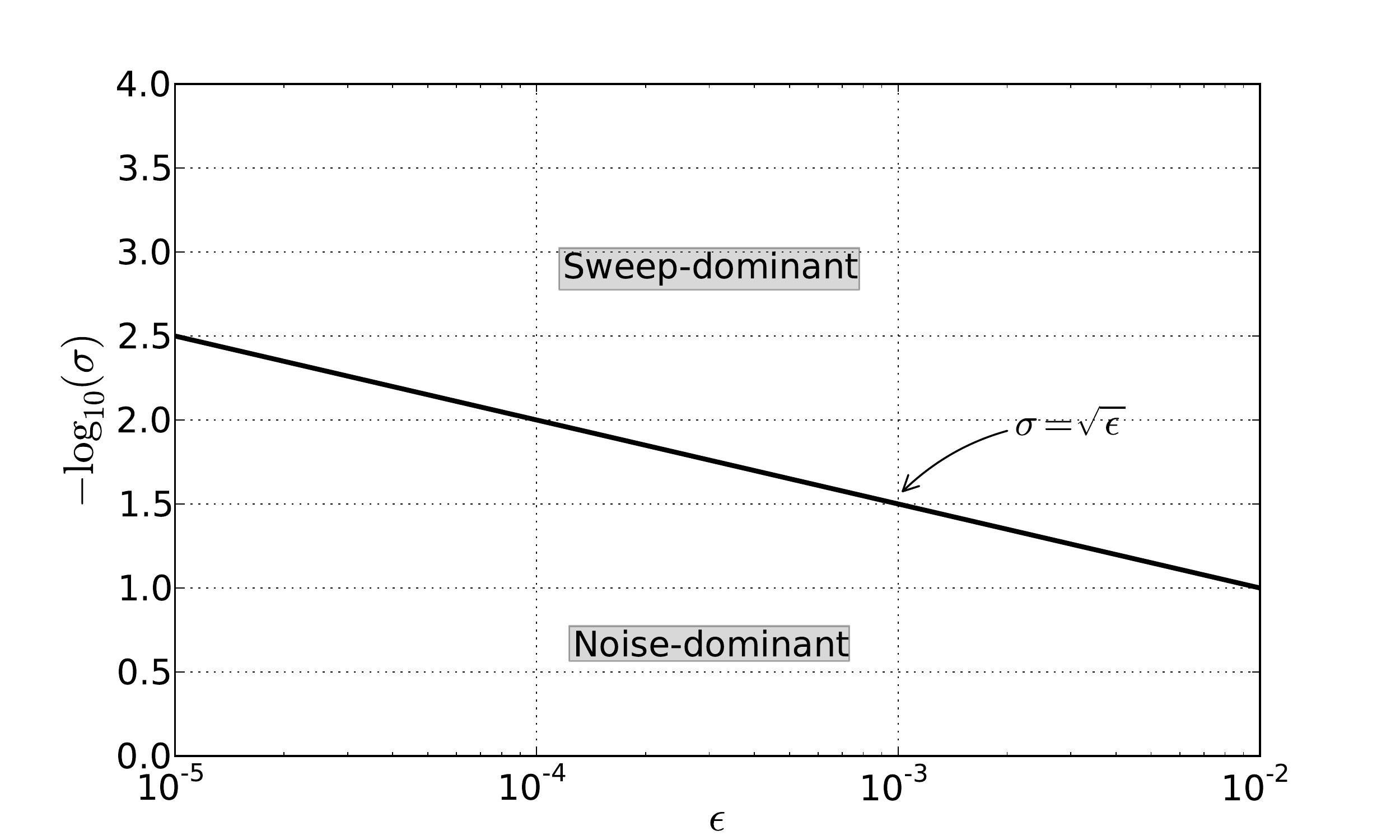}}
\caption{Domain of existence of the deterministic, sweep-dominant and noise-dominant regimes in a plane $(\epsilon, -\log_{10}(\sigma))$. For finite precision cases, $-\log_{10}(\sigma)$ corresponds to the value of the precision.}
\label{fig:SumReg}
\end{figure}

\begin{figure*}[t]
\centering
\includegraphics[width=115mm,keepaspectratio=true]{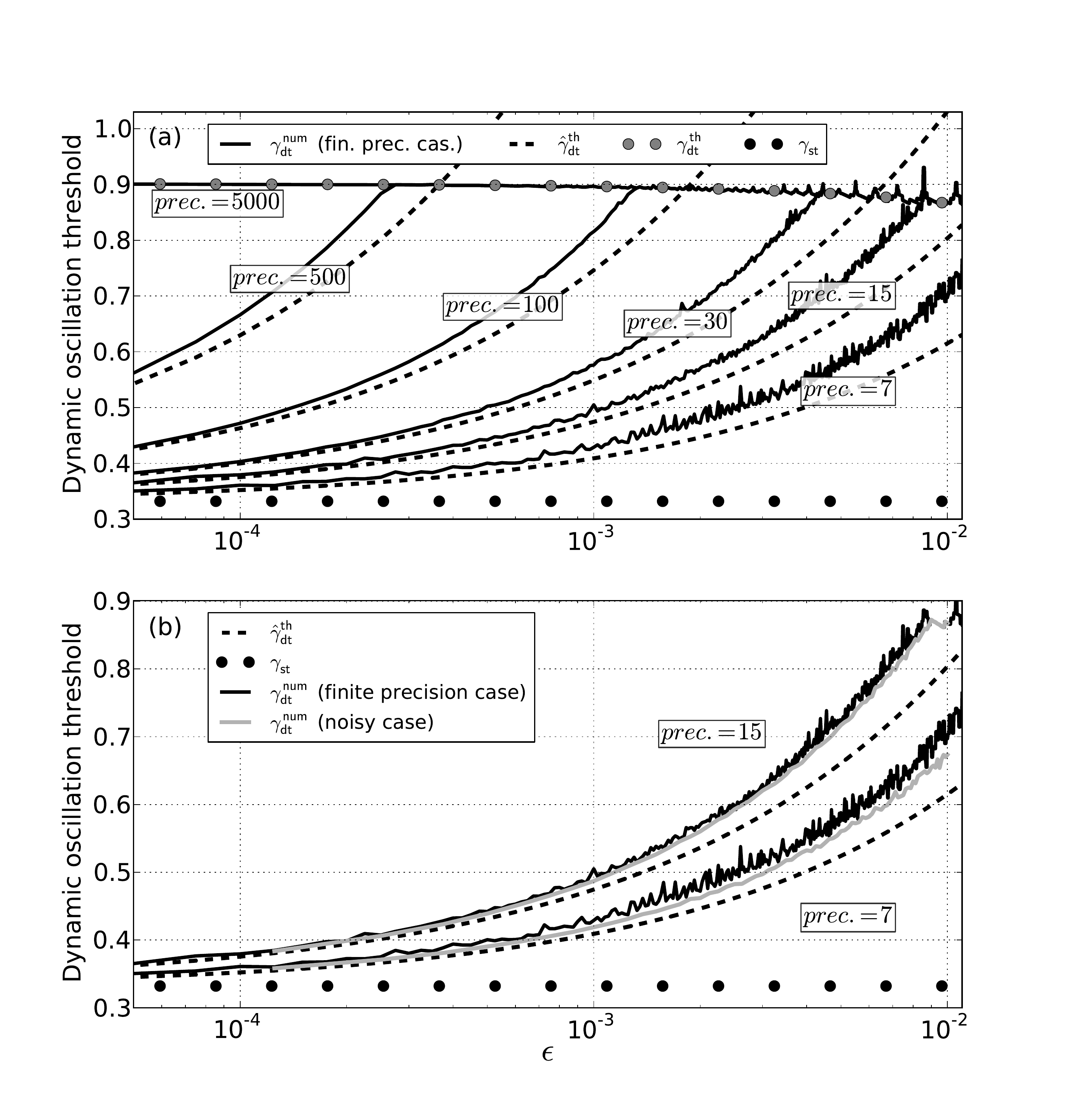}
\caption{Graphical representation of $\gamma_{dt}^{num}$ for different precisions (prec.~= 7, 15, 30, 100, 500 and 5000) with respect to the slope $\epsilon$ and for $\gamma_0 = 0$. Results are also compared to analytical \textit{static} and \textit{dynamic} thresholds: $\gamma_{st}$, $\gamma_{dt}^{th}$ and $\hat{\gamma}_{dt}^{th}$. (a) $\gamma_{dt}^{num}$ and only \textit{finite precision cases} are represented. (b) Both \textit{finite precision cases} and \textit{noisy cases} are represented  for prec. = 7 and 15.}
\label{fig:seuidynnoVSth}
\end{figure*}

By combining equations (\ref{Bn_cont}) and (\ref{I1fin}), $B_n$ is now written as:

\begin{multline}
B_n \approx \\ \frac{\sigma^2}{\epsilon} \int_{\gamma_0+\epsilon}^{\gamma_n+\epsilon} \exp\left(\frac{K}{\epsilon}\left[\left(\gamma_n-\gamma_{st}\right)^2-\left(\gamma'-\gamma_{st}\right)^2\right]\right)d\gamma'\\
= \frac{\sigma^2}{\epsilon}\exp\left(\frac{K}{\epsilon}\left(\gamma_n-\gamma_{st}\right)^2\right)\\ 
\times \underbrace{\int_{\gamma_0+\epsilon}^{\gamma_n+\epsilon} \exp\left(-\frac{K}{\epsilon}\left(\gamma'-\gamma_{st}\right)^2\right)d\gamma'}_{I_2}.
\end{multline}

The function which appears in the integral $I_2$ is a Gaussian function with standard deviation
\begin{equation}
\nu=\sqrt{\frac{\epsilon}{2K}}.
\label{eq:mu}
\end{equation}

Integral $I_2$ is then~\cite{TaOfIntGrad7thEd}:

\begin{equation}
I_2 = \left[\frac{1}{2} \sqrt{\frac{\pi \epsilon}{K}} \, \text{erf}\left(\sqrt{\frac{K}{\epsilon}} \left(\gamma'-\gamma_{st}\right) \right) \right]_{\gamma_0}^{\gamma_n},
\label{eq:I2}
\end{equation}
where erf$(x)$ is the error function. The initial condition $\gamma_0$ is supposed to be much lower than the static threshold $\gamma_{st}$, so that equation (\ref{eq:I2}) can be written:

\begin{equation}
I_2 =\frac{1}{2} \sqrt{\frac{\pi \epsilon}{K}}  \left[ \text{erf}\left(\sqrt{\frac{K}{\epsilon}} \left(\gamma_n-\gamma_{st}\right) \right)+1 \right].
\label{eq:I2simp}
\end{equation}
The dependence on the initial condition $\gamma_0$ is now lost.

Since $\epsilon \ll 1$, for $\gamma_n>\gamma_{st}$ the error function quickly becomes equal to 1 and the integral $I_2$ is simplified to $I_2 =\sqrt{\frac{\pi \epsilon}{K}}$. Finally the expression of $B_n$ is:

\begin{equation}
B_n \approx \frac{\sigma^2}{\sqrt{\epsilon}}\sqrt{\frac{\pi}{K}}\exp\left(\frac{K}{\epsilon}\left(\gamma_n-\gamma_{st}\right)^2\right).
\label{eq:Bfinal}
\end{equation}

From equation (\ref{eq:Bfinal}) it is possible to obtain the expression of $\sqrt{\EV{w_n^2}}\approx\sqrt{B_n}$:

\begin{equation}
\sqrt{\EV{w_n^2}} \approx \sigma \, \epsilon^{-1/4} \left(\frac{\pi}{K}\right)^{1/4} \exp\left(\frac{K}{2\epsilon}\left(\gamma_n-\gamma_{st}\right)^2\right).
\label{eq:sqrtBn}
\end{equation}

The dynamic oscillation threshold $\hat{\gamma}_{dt}^{th}$ is defined \cite{Baesens1991,Baesens1991Noise} as the value of $\gamma_n$ for which the standard deviation $\sqrt{\EV{w_n^2}}$ leaves the neighborhood of the invariant curve. More precisely, the bifurcation occurs when $\sqrt{\EV{w_n^2}}$ becomes equal to the increase rate $\epsilon$, as defined in eq.~(\ref{eq:crit2}). Finally, using equation (\ref{eq:sqrtBn}), we have: 

\begin{equation}
\hat{\gamma}_{dt}^{th} = \gamma_{st}+\sqrt{-\frac{2\epsilon}{K}\ln\left[\left(\frac{\pi}{K}\right)^{1/4}
\frac{\sigma}{\epsilon^{5/4}}\right]},
\label{eq:seuidynno}
\end{equation}
which is the theoretical estimation of the dynamic oscillation threshold of the stochastic systems (\ref{dynsys_pp_no}) (or of the system (\ref{dynsys_pp}) computed using a finite precision) when it evolves in a sweep-dominant regime. The bifurcation delay is a by-product of eq. (\ref{eq:seuidynno}) since it is simply $\hat{\gamma}_{dt}^{th} - \gamma_{st}$.

The method presented in this section is based on a first-order Taylor expansion of the system (\ref{dynsys_pp_no}) around the invariant curve $\phi(\gamma_n)$, leading to the linear system (\ref{dynsys_pplin_no}). Using an asymptotic expansion of the error function it is possible to investigate the behavior of $\sqrt{B_n}$ before $\gamma_n$ enters the neighborhood of the static oscillation threshold $\gamma_{st}$. This study allows to define the domain of validity of this linear approximation, as done by Baesens \cite{Baesens1991,Baesens1991Noise}. This is $\sigma \lesssim \sqrt{\epsilon}$ (more details on obtaining the domain of validity are given in Appendix~\ref{app:B}). Otherwise, if $\sigma\gtrsim \sqrt{\epsilon}$, the orbit of the series $p^+_n$ leaves the neighborhood of invariant curve $\phi(\gamma)$ before the static oscillation threshold is reached. In this case, the linear approximation is no longer valid. This situation is called by Baesens~\cite{Baesens1991,Baesens1991Noise} \textbf{noise-dominant regime} and it is not investigated in the present paper. However,  figure~\ref{fig:SumReg} shows the domain of existence of the different regimes in a plan $[\epsilon \, ; \, -\log_{10}(\sigma)]$. The frontier between deterministic and sweep-dominant regime corresponding to $A_n=B_n$ is determined numerically using the equality $\gamma_{dt}^{th}=\hat{\gamma}_{dt}^{th}$.

The condition $\sigma \lesssim \sqrt{\epsilon}$ is respected in this work since $\sigma=10^{-pr}$ with $7\leq pr \leq 5000$ and $8.10^{-5} \leq \epsilon \leq 10^{-2}$.

\begin{figure}[h!]
\centering
\subfigure[Numerical precision is fixed (prec. = 50).  $\hat{\gamma}_{dt}^{th}$ and $\gamma_{dt}^{num2}$ are computed for $\epsilon=10^{-4}$, $6\cdot 10^{-4}$.]{\includegraphics[width=77mm,keepaspectratio=true]{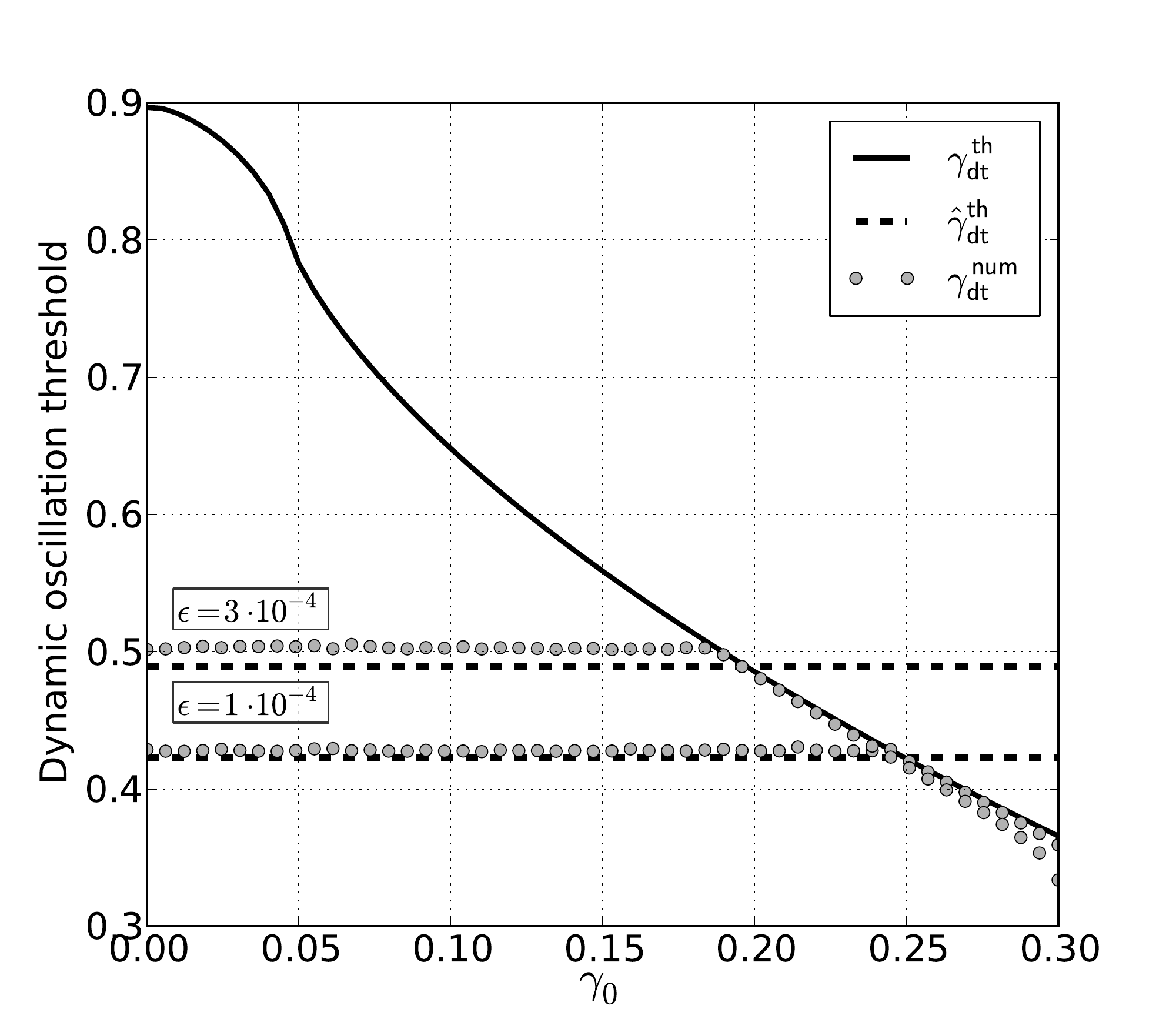}\label{fig:ciprecepsa}}
\subfigure[The increase rate of $\gamma$ is fixed ($\epsilon = 3\cdot 10^{-4}$). $\hat{\gamma}_{dt}^{th}$ and $\gamma_{dt}^{num2}$ are computed for numerical precisions equal to 15 and 100.]{\includegraphics[width=77mm,keepaspectratio=true]{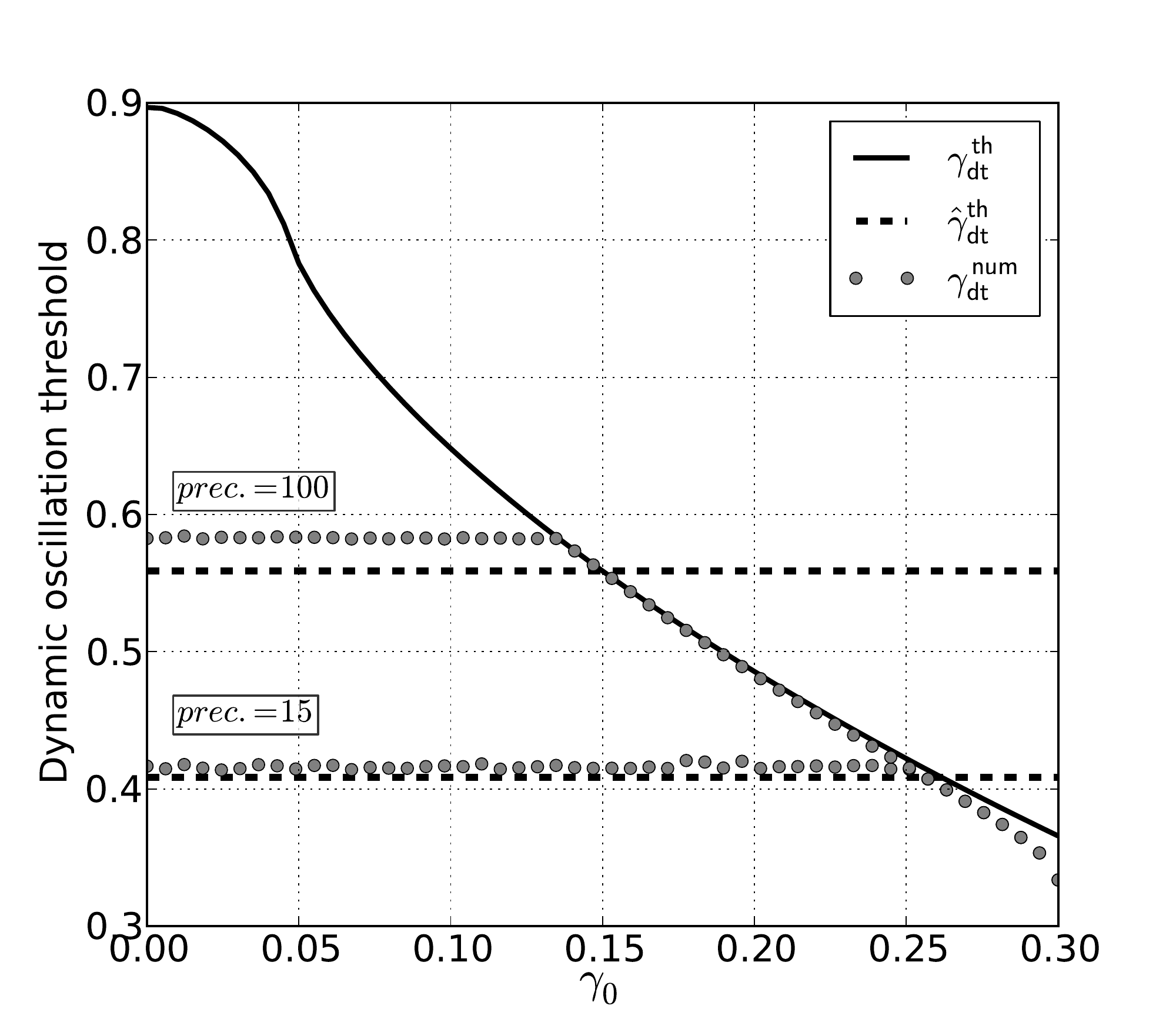}\label{fig:ciprecepsb}}
\caption{Comparison between theoretical prediction of dynamic oscillation threshold (without noise: $\gamma_{dt}^{th}$ and with noise: $\hat{\gamma}_{dt}^{th}$) and the dynamic threshold $\gamma_{dt}^{num}$ computed on numerical simulations for finite precision case. Variable are plotted with respect to the initial condition $\gamma_0$.}
\label{fig:cipreceps}
\end{figure}

\subsection{Discussion}

\begin{figure}[t]
\centering
\includegraphics[width=88mm,keepaspectratio=true]{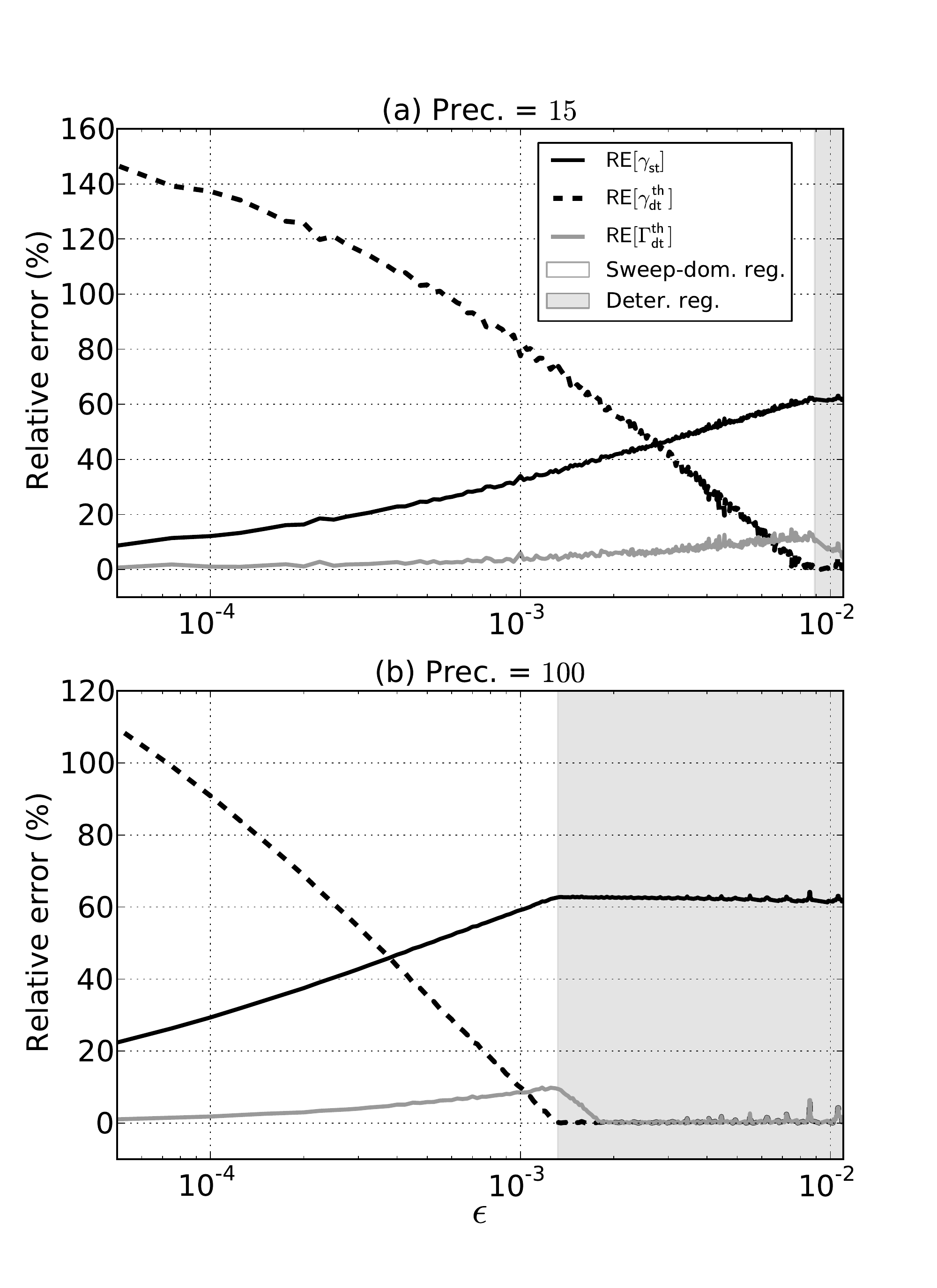}
\caption{Relative errors: $RE\left[\gamma_{st}\right]$, $RE\left[\gamma_{dt}^{th}\right]$ and $RE\left[\Gamma_{dt}^{th}\right]$ for numerical precisions equal to 15 (a) and 100 (b).}
\label{fig:RelErr}
\end{figure}

In figure \ref{fig:seuidynnoVSth}, $\hat{\gamma}_{dt}^{th}$ defined by equation (\ref{eq:seuidynno}) is plotted against the increase rate $\epsilon$. It is compared with $\gamma_{dt}^{num}$ for different values of the precision and for $\gamma_0=0$. In figure \ref{fig:seuidynnoVSth}(a), $\gamma_{dt}^{num}$ is represented  for finite precision cases. The differences between finite precision cases and stochastic cases observed for prec. = 7 and 15 are shown in figure \ref{fig:seuidynnoVSth}(b). The theoretical result $\hat{\gamma}_{dt}^{th}$ provides a good estimation of the dynamic oscillation threshold as long as the system remains in the sweep-dominant regime (with a better estimation when the bifurcation delay is small\footnote{This is an expected result because of the initial assumption of a small bifurcation delay in the presence of noise, leading to first-order Taylor expansions $\gamma_{st}$ in previous calculation (see equation~(\ref{TaExpGamSt})).}). Otherwise, $\gamma_{dt}^{th}$ is a better approximation of $\gamma_{dt}^{num}$, as expected in the deterministic regime.

Figure \ref{fig:cipreceps} shows the comparison between $\hat{\gamma}_{dt}^{th}$ and $\gamma_{dt}^{num}$ (only for finite precision cases) plotted against the initial condition $\gamma_0$. In figure \ref{fig:ciprecepsa}, variables are plotted for several values of $\epsilon$ and for a fixed numerical precision. The opposite is done in figure \ref{fig:ciprecepsb}. As in figure  \ref{fig:seuidynnoVSth}, $\hat{\gamma}_{dt}^{th}$ provides a good estimation of the dynamic oscillation threshold in the sweep-dominant regime, as well as $\gamma_{dt}^{th}$ in the deterministic regime.

Finally, to predict theoretically the dynamic bifurcation threshold $\Gamma_{dt}^{th}$  of the stochastic system (\ref{dynsys_pp_no}) (as well as of the system (\ref{dynsys_pp}) when it is computed with a finite precision) the following procedure is proposed: 

\begin{itemize}
\item compute the theoretical estimation $\hat{\gamma}_{dt}^{th}$ of the stochastic system through eq.~(\ref{eq:seuidynno})
\item compute the theoretical estimation $\gamma_{dt}^{th}$ of the system without noise through eq.~(\ref{dynoscthre_2})
\item if $\hat{\gamma}_{dt}^{th}<\gamma_{dt}^{th}$ the system remains in the ``sweep-dominant regime'' and the dynamic threshold $\Gamma_{dt}^{th}$ is equal to  $\hat{\gamma}_{dt}^{th}$, otherwise the  ``deterministic regime'' is attained and the dynamic threshold $\Gamma_{dt}^{th}$ is equal to $\gamma_{dt}^{th}$.
\end{itemize}

Figure~\ref{fig:RelErr} compares the relative error $RE$ of the three theoretical predictions of the oscillation threshold  ($\gamma_{st}$, $\gamma_{dt}^{th}$ and $\Gamma_{dt}^{th}$)  with respect to $\gamma_{dt}^{num}$, as a percent value:

\begin{equation}
RE[X] = 100 \times \left(\frac{|\gamma_{dt}^{num}-X|}{\gamma_{dt}^{num}}\right),
\end{equation}
where $X$ takes successively the values of $\gamma_{st}$, $\gamma_{dt}^{th}$ and $\Gamma_{dt}^{th}$.

For standard double-precision (fig.~\ref{fig:RelErr}(a), prec.=15), the sweep-dominant regime is prevalent throughout most of the range of increase-rates studied in this article. Higher precisions (for instance prec.=100) imply the appearence of the deterministic-regime for lower increase-rates.  In this case, $\Gamma_{dt}^{th}$ provides a better estimation of the oscillation threshold of the clarinet with a linearly increasing blowing pressure. Indeed, in situations represented in figure \ref{fig:RelErr}, $RE\left[\Gamma_{dt}^{th}\right]$ never exceeds $15\%$ while $RE\left[\gamma_{st}\right]$ and $RE\left[\gamma_{dt}^{th}\right]$ can reach $60\%$ and $145\%$ respectively.  At slightly lower values of $\epsilon$ than the limit between the two regimes, $\gamma_{dt}^{th}$ still provides a better estimation of $\gamma_{dt}^{num}$ than $\Gamma_{dt}^{th}$, a situation that occurs for all values of the precision, according to figure \ref{fig:seuidynnoVSth}.

\section{Conclusion}

In many situations, the finite precision used in numerical simulations of the clarinet system does not produce major errors in the final results that are sought. Such is the case, for instance, when estimating the amplitudes for a given regime. 

However, when slowly increasing one of the control parameters, the distances between the state of the system and the invariant curve can become smaller than the round-off errors of the calculation, with dramatic effects on the time required to trigger an oscillation. In these cases, the inclusion of a stochastic variable in the theory allows to correctly estimate the threshold observed in simulations, which lies between the static and dynamic thresholds found for precise cases.

As a final remark, the present theoretical study is probably not restricted to describe numerical simulations. Indeed, the noise level $\sigma$ measured in  an artificially blown instrument is typically of the order of magnitude of $10^{-3}$. The domain of validity of the results: $\sigma \lesssim \sqrt{\epsilon}$ suggests that the comparison with experiment using blowing pressure with increase rates $\epsilon>10^{-6}$ (typically for usual clarinets that corresponds to $\approx 5$Pa/s),  seems to be possible although the noise level usually increases with the pressure applied to the instrument. 

It is known that the static oscillation of the clarinet is difficult to measure by increasing, even slowly, the mouth pressure. The phenomenon of dynamic bifurcation is a possible reason. If that were proven experimentally, we could imagine to inverse the equation~(\ref{eq:seuidynno}) to deduce the static threshold from the measurement of the noise level, the increase rate of the blowing pressure and the dynamic threshold.

\subsection*{Acknowledgements}
This work was done within the framework of the project SDNS-AIMV "Syst\`{e}mes Dynamiques Non-Stationnaires - Application aux Instruments \`{a} Vent" financed by \emph{Agence Nationale de la Recherche} (ANR).

\appendix

\section{Limit of the linear calculation}
\label{app:B}

The method presented in Section \ref{sec:SeDynNoTh} is based on a first-order Taylor expansion of the system (\ref{dynsys_pp_no}) around the invariant curve $\phi(\gamma_n)$ leading to define the linear system (\ref{dynsys_pplin_no}). Following Baesens  \cite{Baesens1991,Baesens1991Noise}, we give here the upper bound of the domain of validity of this linear approximation.

Using equations (\ref{Bn_cont}) and (\ref{eq:I2simp}), the expression $B_n$ is given by:

\begin{multline}
B_n \; =
\frac{\sigma^2}{2}\sqrt{\frac{\pi}{\epsilon \; K}} \exp\left(\frac{K}{\epsilon}\left(\gamma_n-\gamma_{st}\right)^2\right) 
\\
\times \left[ \text{erf}\left(\sqrt{\frac{K}{\epsilon}} \left(\gamma_n-\gamma_{st}\right) \right)+1 \right].
\label{eq:B1}
\end{multline}

We investigate the behavior of $\EV{w_n^2}$ before $\gamma_n$ enters in the neighborhood of the static oscillation threshold $\gamma_{st}$. More precisely, we compute an approximate expression of $\EV{w_n^2}$ when $\gamma_n<\gamma_{st}-\nu$, where $\nu$ is defined by equation (\ref{eq:mu}). To do this, the error function in equation (\ref{eq:B1}) is expanded in a first-order asymptotic series~\cite{HandbookMathAbra} (the asymptotic expansion of the error function erf$(x)$ for large negative $x$ is recalled in Appendix \ref{ann:asyexperf}):

\begin{multline}
B_n  \; =
\frac{\sigma^2}{2}\sqrt{\frac{\pi}{\epsilon \; K}} \exp\left(\frac{K}{\epsilon}\left(\gamma_n-\gamma_{st}\right)^2\right) 
\\
\times \left[-1-\frac{\exp\left(-\frac{K}{\epsilon}\left(\gamma_n-\gamma_{st}\right)^2\right)}{\sqrt{\frac{K \pi }{\epsilon}}\left(\gamma_n-\gamma_{st}\right)}+1\right],
\label{eq:B2}
\end{multline}
which is simplified in:

\begin{equation}
B_n \; = -\frac{\sigma^2}{2K\left(\gamma_n-\gamma_{st}\right)}.
\label{eq:B3}
\end{equation}

Using the explicit form of $\gamma_n$, solution of equation (\ref{dynsys_pp_b}):

\begin{equation}
\gamma_n = \epsilon \, n + \gamma_0,
\label{eq:B4}
\end{equation}
and (\ref{eq:B3}), we have:

\begin{equation}
\sqrt{B_n} \; = \frac{\sigma}{\sqrt{2K\epsilon}}\frac{1}{\sqrt{n_{st}-n}},
\label{eq:B5}
\end{equation}
where $n_{st}$ is the iteration for which $\gamma_{st}$ is reached.

Equation (\ref{eq:B5}) means that when $\gamma_n<\gamma_{st}-\nu$, the standard deviation $\sqrt{\EV{w_n^2}}\approx\sqrt{B_n}$ increases with the time (i.e. with $n$) like $1/\sqrt{n_{st}-n}$ to order $\sigma/\sqrt{\epsilon}$, and therefore remains small if $\sigma\ll \sqrt{\epsilon}$. Otherwise, if $\sigma\gtrsim \sqrt{\epsilon}$, the orbit of the series $p^+_n$ leaves the neighborhood of invariant curve $\phi(\gamma)$ before the static oscillation threshold is reached. In this case, linear calculation made in Section \ref{sec:SeDynNoTh} is no longer valid.

\section{Asymptotic expansion of error function}
\label{ann:asyexperf}

The asymptotic expansion of the error function erf$(x)$ for large negative $x$ ($x\rightarrow -\infty$) is~\cite{HandbookMathAbra}:

\begin{multline}
\text{erf}(x) \approx -1 - \frac{\exp\left(-x^2\right)}{\sqrt{\pi}x}
\\
\times \left(1+\sum_{m=1}^{+\infty}\left(-1\right)^m\frac{1\cdot3 \, \ldots \, (2m-1)}{\left(2x^2\right)^m}\right)
\end{multline}


\end{document}